%% file: chatlas.tex
\documentclass{article}

\PassOptionsToPackage{numbers, compress}{natbib}
\usepackage[preprint]{neurips_2024}
\makeatletter
\renewcommand{\@noticestring}{}
\makeatother

\usepackage[utf8]{inputenc} 
\usepackage[T1]{fontenc}    
\usepackage{hyperref}       
\usepackage{url}            
\usepackage{booktabs}       
\usepackage{amsfonts}       
\usepackage{nicefrac}       
\usepackage{microtype}      
\usepackage{xcolor}         
\usepackage{array}
\usepackage{graphicx}
\usepackage{amsmath}
\usepackage[capitalise]{cleveref}
\usepackage{algorithm}    
\usepackage{algpseudocode}  
\usepackage{xspace}
\usepackage{placeins}
\usepackage[colorinlistoftodos]{todonotes}
\usepackage{tikz}
\usetikzlibrary{arrows.meta, positioning, calc}
\usepackage{fontawesome5}
\usepackage{subcaption}
\usepackage{comment}

\title{ch\textsc{ATLAS}: A Retrieval-Augmented Conversational Assistant for the ATLAS Experiment}

\author{%
\\
Joe Egan\textsuperscript{1}, 
Ben Elliot\textsuperscript{1}, 
Samuel Van Stroud\textsuperscript{1}, 
Alex Sopio\textsuperscript{2},
Ki Ryeong Park\textsuperscript{3}, \\
Daniele Dal Santo\textsuperscript{4},
Zhaohui Li\textsuperscript{1},
Runze Li\textsuperscript{5},
Jeremy Couthures\textsuperscript{6},
Joe George\textsuperscript{1},
Cary Randazzo\textsuperscript{7},\\
Juerg Beringer\textsuperscript{8},
Daniel Murnane\textsuperscript{8,9,*},
Gabriel Facini\textsuperscript{1,*}, \\
for the ATLAS Software and Computing group.
\\
\\
\textsuperscript{1}\textit{Centre for Data Intensive Science and Industry, University College London, London, UK} \\
\textsuperscript{2}\textit{University of Edinburgh, Edinburgh, UK} \\
\textsuperscript{3}\textit{Nevis Laboratory, Columbia University, NY, USA} \\
\textsuperscript{4}\textit{Universität Bern, Bern, Switzerland} \\
\textsuperscript{5}\textit{Yale University, New Haven, CT, USA} \\
\textsuperscript{6}\textit{Laboratoire d’Annecy de Physique des Particules, Annecy, France} \\
\textsuperscript{7}\textit{Louisiana Tech University, Ruston, LA, USA} \\
\textsuperscript{8}\textit{Lawrence Berkeley National Laboratory, Berkeley, CA, USA} \\
\textsuperscript{9}\textit{University of Copenhagen, Copenhagen, Denmark} \\
}

\newcommand{\ours}{\texttt{ch}\textsc{atlas}\xspace}

\begin{document}


\maketitle
\begingroup
\renewcommand{\thefootnote}{*}
\footnotetext[0]{Corresponding authors: \href{mailto:daniel.murnane@nbi.ku.dk}{daniel.murnane@nbi.ku.dk}; \href{mailto:gabriel.facini@ucl.ac.uk}{gabriel.facini@ucl.ac.uk}.}
\endgroup

\begin{abstract}
Large scientific collaborations produce extensive internal documentation spread across heterogeneous systems, making it difficult for collaborators to locate and reuse existing knowledge. We present \ours, a deployed Retrieval-Augmented Generation (RAG) assistant developed for the ATLAS experiment at CERN. The system provides conversational access to more than 30,000 documents from several collaboration resources, material largely unavailable to general-purpose language models, using hybrid dense--sparse retrieval deployed on CERN infrastructure.
A domain-specific benchmark built from automatically generated question--answer pairs is used to evaluate retrieval and generation performance and to guide design choices. We study the selection and domain-specific fine-tuning of an embedding model, contextual retrieval, ranking strategies, and generator choice. Fine-tuning combined with contextual retrieval more than doubles how often the correct document is returned as the top result on the most technically challenging documentation source. The deployed system also includes intent routing, safety handling, Mattermost integration, and an MCP server for future agentic workflows.
\ours demonstrates that domain-adapted RAG can provide source-attributed, domain-specific access to collaboration knowledge, and offers a practical foundation for scientific collaborations and other organisations that rely on accurate, up-to-date retrieval from proprietary documentation.

\end{abstract}

\input{sections/1.introduction}
\input{sections/2.related}

\input{sections/3.atlasdocs}
\input{sections/4.benchmark}
\input{sections/5.core}
\input{sections/6.design}
\input{sections/7.conclusions}


\FloatBarrier

\clearpage

\section*{Acknowledgments}
The development of \ours is part of the offline software research and development programme of the ATLAS Collaboration, and we thank the collaboration for its support and cooperation. We are especially grateful to Zachary Marshall, Edward Moyse, and David South for supporting the project and coordinating the ATLAS review, and to David Rousseau, Michael Kagan, Edward Moyse, Xiangyang Ju, and Max Swiatlowski for comments during that review. We thank Max Hart for encouraging the development of the first prototype and for useful early technical advice, and Jason Oliver for fruitful discussions and a shared commitment to improving search across ATLAS documentation. We further thank Subhashis Suara and the CERN IT Department for their work on the custom ATLAS TWiki back end.
This work is funded in part by the UK’s Science and Technology Facilities Council via University
College London’s Centre for Doctoral Training in Data Intensive Science and by departmental and industry contributions. DM is supported by the Danish Data Science Academy, which is funded by the Novo Nordisk
Foundation (NNF21SA0069429). 

\subsection*{Author Contributions (CRediT)}

\textbf{Conceptualization:} D.M., G.F. \\
\textbf{Methodology:} J.E., B.E., S.V.S., A.S., R.L., D.D.S., Z.L., J.C., J.B., D.M., G.F. \\ 
\textbf{Software:} All authors. \\
\textbf{Validation:} J.E., B.E., S.V.S., J.B., D.M. \\
\textbf{Investigation:} J.E., B.E., S.V.S., A.S., R.L., Z.L., J.C., J.B., D.M., G.F. \\
\textbf{Formal analysis:} J.E., B.E., S.V.S., Z.L., K.R.P, J.B., D.M. \\
\textbf{Data Curation:} J.E., B.E., S.V.S., R.L., D.D.S., C.R., G.F. \\
\textbf{Writing -- Original Draft:} J.E., J.B., G.F. \\
\textbf{Writing -- Review \& Editing:} J.E., B.E., S.V.S, A.S., K.R.P, Z.L., J.B., D.M., G.F. \\
\textbf{Supervision:} J.E., J.B., D.M., G.F. \\
\textbf{Project Administration:} J.B., D.M., G.F.


\newpage

\bibliography{chatlas}

\newpage


\appendix
\crefalias{section}{appendix}
\crefalias{subsection}{appendix}
\crefname{appendix}{Appendix}{Appendices}
\Crefname{appendix}{Appendix}{Appendices}

\input{sections/8.appendix}


\newpage

\end{document}

%% file: sections/1.introduction.tex
\section{Introduction}
\label{sec:intro}

High energy physics (HEP) experiments such as those on CERN's Large Hadron Collider operate at an unprecedented scale, both in terms of the data produced and the size of the collaboration. The ATLAS Collaboration is one of the largest collaborative efforts ever undertaken in science, numbering about 6,000 members from more than 170 institutions worldwide~\cite{atlas_membership}, and the experiment generates $\mathcal{O}(50)$ petabytes of collision data every year. Supporting this activity is a vast and diverse information ecosystem, including technical reports, internal notes, meeting minutes, presentation slides and recordings, software documentation, monitoring dashboards, and wiki pages. An example is ATLAS TWiki, containing over 35,000 pages written over more than two decades by hundreds of contributors. While this abundance of documentation reflects the collaborative strength of the experiment, it also creates a significant challenge. Information is scattered across many sources, written in many voices, inconsistently maintained, and difficult to search effectively. As a result, even experienced collaborators often struggle to locate the right information quickly, and new members face a steep learning curve in their first years on the experiment.

The information retrieval challenge within the scientific community and high energy physics has been recognized for more than two decades. As early as 2003, an ATLAS white paper highlighted the ``opportunities for use and development of collaborative tools'' and warned of the risks posed by fragmented and un-curated documentation~\cite{goldfarb2003atlas}. Two decades later, these concerns remain pressing. The 2024 ``Big ATLAS Survey'' identified documentation and searchability as collaboration-wide concerns, prompting efforts to improve the accessibility and organisation of ATLAS knowledge~\cite{atlas2025survey,atlasBigSurveyCourier2026}. While ATLAS has abundant documentation, its dispersion across many systems and styles makes it hard to access and use effectively.

The costs of this fragmentation are high. New collaborators, often PhD students, face a steep ramp-up time to locate the information they need. Established members waste effort re-discovering knowledge or rely on informal channels via personal connections. The collaboration as a whole risks reduced efficiency, creativity, and impact. A unified assistant that can surface relevant information quickly and reliably could flatten the learning curve, free up time for more complex tasks, and improve the collective output of the collaboration. However, such an assistant can only be as good as the documentation it draws upon. The accuracy, completeness, and timeliness of the underlying sources ultimately bound the quality of its responses.

Recent advances in artificial intelligence, particularly Large Language Models (LLMs), provide new opportunities to support scientific work by automating and accelerating information-intensive tasks~\cite{LLMsInScience,Hu2025ASO}. LLMs are neural sequence models based on the attention mechanism using the transformer architecture~\cite{vaswani2017attention}. Popularised by general-purpose conversational assistants such as OpenAI's ChatGPT, they have been shown to perform effectively across a range of domains. Notable examples include code generation~\cite{llmcodegen2025survey}, as in Copilot/Codex~\cite{chen2021evaluating}, summarisation and synthesis of information across multiple documents~\cite{zhang2024benchmarking}, and the provision of context-sensitive answers to natural language queries by integrating evidence from several passages. 
However, there are well-recognised limitations when applying LLMs to domain-specific tasks. A central issue is that such models are trained on a fixed snapshot of predominantly public data. As a consequence, they do not incorporate the internal documentation generated within large scientific collaborations. For ATLAS, the majority of relevant material, including technical notes, meeting records, and software documentation, is internal and therefore inaccessible to commercial LLMs during training. Retrieval-Augmented Generation (RAG)~\cite{lewis2020rag} addresses this limitation by coupling a generative model to an \textit{external retriever}, which identifies relevant internal documents at query time and supplies them as additional input, or \textit{conditioning context}, for response generation. 
In this paper we present \ours, a multi-turn conversational system that applies the RAG paradigm to the ATLAS knowledge base. \ours is designed to help collaborators discover and understand documentation. It is not intended as a software-coding assistant, although code-aware extensions are discussed as a direction for future work in \cref{sec:conclusions}.

This paper is organised as follows. \Cref{sec:rag} introduces RAG techniques and positions \ours relative to related work in science and industry.
\Cref{sec:atlasdocs} includes an overview of ATLAS documentation. 
\Cref{sec:evaluation} introduces the benchmarking techniques developed for \ours, while \cref{sec:core} describes the current implementation of \ours. 
Whenever possible, design choices were made to improve the benchmark results and a set of supporting studies is discussed in \cref{sec:decisions}.
Finally, a concluding discussion and direction of future work is given in \cref{sec:conclusions}.


%% file: sections/2.related.tex
\section{RAG Background and Related Work}
\label{sec:rag}

\subsection{RAG Architecture and the \ours Pipeline}

Retrieval-Augmented Generation (RAG) is a framework that combines a retriever with a generator. The retriever selects relevant passages from a corpus. The generator, typically an LLM, produces an answer to the query using the retrieved passages as context. This architecture addresses two key weaknesses of LLMs trained only on fixed corpora. It gives the model access to up-to-date information and reduces the tendency to hallucinate unsupported claims~\cite{lewis2020rag,Gao2023RetrievalAugmentedGF}. From a user perspective, interaction with a RAG system is indistinguishable from a direct conversation with an LLM as queries are typed in natural language and responses are returned in the same conversational style. 
Source retrieval and ranking steps are carried out automatically, so the user does not need to manage them, but the underlying sources can still be shown alongside the generated answer. This combination preserves the fluency of an LLM while ensuring that responses are grounded in a controlled corpus of documents. Such grounding is particularly valuable in scientific collaborations, 
where the majority of documentation is restricted to collaboration members and serves as the definitive reference from detector operations to data analysis. Making this material accessible through RAG ensures that responses remain both trustworthy and verifiable~\cite{Barman:2025wfb}. 
One additional advantage of semantic interaction is the potential for non-English queries to be used to search the predominantly English-language corpus. This is particularly relevant for a global collaboration such as ATLAS, where non-native English speakers may face additional barriers when navigating technical documentation.

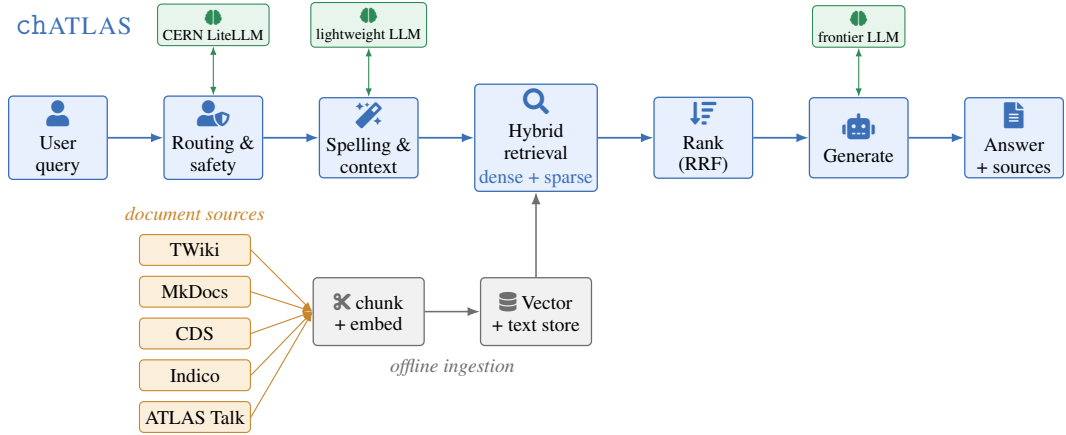
\begin{figure}
    \centering
    \definecolor{onfill}{HTML}{E8F0FE}%
    \definecolor{ondraw}{HTML}{3B6FB6}%
    \definecolor{stfill}{HTML}{F1F1F1}%
    \definecolor{stdraw}{HTML}{6E6E6E}%
    \definecolor{srcfill}{HTML}{FBEEDB}%
    \definecolor{srcdraw}{HTML}{C8862B}%
    \newcommand{\picon}[1]{{\large\textcolor{ondraw}{\faIcon{#1}}}}%
    \newcommand{\spicon}[1]{{\textcolor{stdraw}{\faIcon{#1}}}}%
    \definecolor{mfill}{HTML}{E7F5EC}%
    \definecolor{mdraw}{HTML}{2E8B57}%
    \newcommand{\micon}[1]{{\textcolor{mdraw}{\faIcon{#1}}}}%
    \resizebox{\linewidth}{!}{%
    \begin{tikzpicture}[
      font=\small, >=Latex, node distance=6mm and 9mm,
      stage/.style={draw=ondraw, line width=.6pt, rounded corners=2pt, fill=onfill,
                    align=center, minimum height=13mm, minimum width=16mm, inner sep=2pt},
      store/.style={draw=stdraw, line width=.6pt, rounded corners=2pt, fill=stfill,
                    align=center, minimum height=11mm, minimum width=18mm, inner sep=2pt},
      src/.style={draw=srcdraw, line width=.5pt, rounded corners=2pt, fill=srcfill,
                  align=center, font=\footnotesize, minimum height=5mm, minimum width=18mm, inner sep=1.5pt},
      ar/.style={-{Latex[length=2mm]}, line width=.8pt, draw=ondraw},
      arg/.style={-{Latex[length=2mm]}, line width=.7pt, draw=stdraw},
      ars/.style={-{Latex[length=1.6mm]}, line width=.5pt, draw=srcdraw},
      model/.style={draw=mdraw, line width=.5pt, rounded corners=2pt, fill=mfill,
                    align=center, font=\scriptsize, inner sep=2.5pt, minimum width=15mm},
      arm/.style={{Latex[length=1.4mm]}-{Latex[length=1.4mm]}, line width=.5pt, draw=mdraw}]
      \node[stage] (q) {\picon{user}\\[2pt]User\\query};
      \node[stage, right=of q]  (rt) {\picon{user-shield}\\[2pt]Routing \&\\safety};
      \node[stage, right=of rt] (pp) {\picon{magic}\\[2pt]Spelling \&\\context};
      \node[stage, right=of pp] (rg) {\picon{search}\\[2pt]Hybrid\\retrieval\\[1pt]{\footnotesize\textcolor{ondraw}{dense + sparse}}};
      \node[stage, right=of rg] (rk) {\picon{sort-amount-down}\\[2pt]Rank\\(RRF)};
      \node[stage, right=of rk] (gn) {\picon{robot}\\[2pt]Generate};
      \node[stage, right=of gn] (an) {\picon{file-alt}\\[2pt]Answer\\+ sources};
      \foreach \a/\b in {q/rt, rt/pp, pp/rg, rg/rk, rk/gn, gn/an} \draw[ar] (\a)--(\b);
      \node[model, above=8mm of rt] (mrt) {\micon{brain}\\CERN LiteLLM};
      \node[model, above=8mm of pp] (mpp) {\micon{brain}\\lightweight LLM};
      \node[model, above=8mm of gn] (mgn) {\micon{brain}\\frontier LLM};
      \draw[arm] (rt) -- (mrt);
      \draw[arm] (pp) -- (mpp);
      \draw[arm] (gn) -- (mgn);
      \node[anchor=west, font=\Large, text=ondraw] at (q.west |- mrt) {\ours};
      \node[store, below=16mm of pp.south, anchor=north] (ce) {\spicon{cut}~chunk\\+ embed};
      \node[store] (vs) at (rg |- ce) {\spicon{database}~Vector\\+ text store};
      \draw[arg] (ce)--(vs);
      \draw[arg] (vs.north) -- (rg.south);
      \node[src, left=10mm of ce, yshift=10mm] (s1) {TWiki};
      \node[src, below=1.6mm of s1] (s2) {MkDocs};
      \node[src, below=1.6mm of s2] (s3) {CDS};
      \node[src, below=1.6mm of s3] (s4) {Indico};
      \node[src, below=1.6mm of s4] (s5) {ATLAS Talk};
      \foreach \s in {s1,s2,s3,s4,s5} \draw[ars] (\s.east) -- (ce.west);
      \node[font=\footnotesize\itshape, text=srcdraw, above=1mm of s1] {document sources};
      \node[font=\footnotesize\itshape, text=stdraw] at ($(ce.south)!0.5!(vs.south)+(0,-3.5mm)$) {offline ingestion};
    \end{tikzpicture}}
    \caption{Schematic overview of the \ours pipeline. Online (top), each user query passes through intent routing and safety screening, spelling correction and contextualisation, hybrid (dense and sparse) retrieval, rank fusion, and generation, returning an answer together with its sources. Stages that invoke a language model are annotated in green. Routing and safety use a CERN-hosted model served through the LiteLLM gateway, query preparation uses a separate lightweight model, and generation uses a frontier model. Offline (bottom), documents from multiple ATLAS sources are chunked, embedded, and stored in the combined vector and text store that retrieval queries.}
\label{fig:rag_pipeline}
\end{figure}

The \ours RAG pipeline follows standard stages. Source documents are organised and \textit{chunked} into retrievable units, often with additional metadata. These units are embedded into semantic vectors using sentence encoders and stored in a database. User queries may pass through intent routing, safety checks, and reformulation or expansion. Candidate documents are then retrieved, ranked, and passed to the generator, which produces a factually-grounded response as shown in \cref{fig:rag_pipeline}. Variants extend this loop with self-correction, iterative retrieval, or agentic decision-making. In the following subsections we review each of these components and the relevant literature, framing them in the context of the ATLAS documentation ecosystem.

\subsection{Components of a RAG System}

\paragraph{Data organisation.}
Because an embedding model represents an entire passage as a single fixed-length vector (described below), long documents that span several topics lose information when encoded as a whole. Splitting documents into smaller units for retrieval can mitigate this loss.
The strategies used vary from fixed-length windows, sliding context, or semantically segmented chunks and can have sizeable effects on retrieval quality~\cite{Zhong2024MixofGranularityOT}. Attaching metadata (e.g.\ author, date, or the relevant ATLAS working group---an organisational unit responsible for a particular detector, physics, software, or computing area) provides additional handles for filtering and ranking. A popular strategy that balances granularity and context is the \emph{parent--child structure}~\cite{llamaindex-pc}, in which a full document or major section is treated as the parent, and the retrievable chunks derived from it are linked as children. This design allows metadata and provenance to be inherited from the parent while retrieval and ranking operate at the level of the children. During generation, the parent document can be passed to the LLM to provide additional context, preserving both traceability and interpretability. Extensions such as GraphRAG~\cite{graphrag2025} go further by constructing explicit graphs of relationships, linking not only chunks via citations or hyperlinks but also entities and concepts across documents, enabling retrieval that exploits both document structure and domain-specific connections.

\paragraph{Embeddings.}
Retrieval operates on \emph{embeddings}, which are dense vectors that place semantically similar text close together in a shared vector space. Text is first split into \emph{tokens}, sub-word units determined by the tokenizer of the model (for English, one token typically corresponds to three to four characters), and a transformer maps each token to its own vector. For retrieval, these per-token vectors are pooled into a single fixed-length \emph{sentence} (or \emph{passage}) embedding representing the whole chunk, with the same model embedding both document chunks and user queries. General-purpose encoders (e.g.\ Sentence-BERT~\cite{reimers2019sentencebert}) work well, but domain-adapted embeddings can be obtained via fine-tuning on ATLAS-specific material. Smaller models may be advantageous for efficiency and local inference, whereas larger models typically offer richer representations at higher computational cost.

\paragraph{Query manipulation.}
Retrieving the correct documents from the database can be limited by the format of the user query. For example, queries that are too short, use synonyms not present in the corpus, or fail to match the phrasing of stored documents lead to poor search results. Query manipulation techniques address this by reformulating the query before a search is performed. Examples include query expansion (adding related terms), prompt reformulation (re-phrasing the question into a form closer to the target documents), and synthetic document generation such as Query2doc~\cite{Wang2023Query2docQE} or Hypothetical Document Expansion (HyDE)~\cite{hyde2022}, where an LLM first generates relevant information or a plausible answer that is then included in the retrieval query.

\paragraph{Search.}
In a typical dense retrieval pipeline, the user query is first embedded into the same vector space as the document chunks using the chosen encoder model. Candidate documents are then identified by comparing the query embedding $\mathbf{q}$ to those of all chunks $\mathbf{d}_i$, usually with cosine similarity
\begin{equation}
\mathrm{sim}_{\cos}(\mathbf{q}, \mathbf{d}_i) = \frac{\mathbf{q} \cdot \mathbf{d}_i}{\|\mathbf{q}\| \, \|\mathbf{d}_i\|},
\end{equation}
or inner-product distance $\mathrm{sim}_{\mathrm{dot}}(\mathbf{q}, \mathbf{d}_i) = \mathbf{q} \cdot \mathbf{d}_i$ as the similarity metric. 
This retrieval method exploits the property of high-dimensional vector spaces that two randomly chosen vectors are vanishingly unlikely to point in the same direction.
Here, ``dense'' refers to the use of compact, continuous embeddings (typically a few hundred dimensions) learned by neural networks. As corpora can contain millions of chunks, efficient approximate nearest-neighbour algorithms and vector stores (e.g.\ FAISS~\cite{johnson2017,douze2024faiss}) are used to keep retrieval latency low. Alternatively, exact similarity search can be performed at higher computational cost.  
Classical ``sparse'' retrieval approaches such as BM25~\cite{robertson94,robertson09} rank documents by using term-frequency (TF) vectors to match query terms and weight them according to inverse document frequency (IDF), which reduces the influence of common words and emphasises rarer, more informative terms. 
Hybrid systems often combine both dense (embedding-based) and sparse (keyword-based) methods, improving recall and robustness by covering cases where one method fails and using coincidence between the search types as a sign of high confidence in the relevance of a document.

\paragraph{Ranking.}
After retrieval, candidate passages must be ranked so that the most relevant evidence is placed within the generator's finite context window. Using increasingly large contexts is both costly and susceptible to context rot, a systematic degradation in model reliability and reasoning accuracy as input length increases~\cite{hong2025contextrot}. This phenomenon encompasses the well-documented ``lost in the middle'' effect~\cite{liu2023lost,hsieh2024ruler}, where important details are diluted or ignored when buried among irrelevant material.
\textit{Ranking} is a group of algorithms used to combine results from multiple retrievers in an ordered way. This is especially useful when combining results from text and semantic search, where the scoring metrics can have very different scales. This also provides a method for upweighting results returned from both retrievers, which are more likely to be relevant.
Ranking maximises relevant signal per token, which is often the dominant cost driver in production RAG systems. 
A simple cut on the retrieval metric or a limited number of retrieved documents can be used when only one type of search is performed. Reciprocal Rank Fusion (RRF)~\cite{rrf2009} is a simple but effective way to combine rankings from multiple retrievers with
\begin{equation}
\mathrm{RRF}(d) = \sum_{r \in R} \frac{1}{k + \mathrm{rank}_{r}(d)},
\label{eq:rrf}
\end{equation}
where $R$ is the set of individual rankers, $\mathrm{rank}_{r}(d)$ is the rank position of document $d$ in ranker $r$, and $k$ is a constant (typically $k = 60$) that dampens the contribution of lower-ranked items.
With a higher computational cost, cross-encoder re-ranking~\cite{nogueira2019,nogueira2020,gao2021} uses a bi-directional transformer to score query--document relevance with high accuracy.

\paragraph{Advanced RAG.}
Standard RAG architectures operate in a single direction. A query triggers retrieval, and the retrieved passages are passed to the generator for response synthesis. More advanced variants introduce decision points that allow the model to control this process dynamically. The decision points can either be implemented as a static logic chain or as an \emph{agent}. In this setting, the agent is an LLM configured to decide whether to search, generate, or call an external tool. The retrieval--generation loop can also be enhanced with self-correction or iterative retrieval.
Agentic frameworks such as ReAct~\cite{yao2023react}, Corrective RAG (CRAG)~\cite{crag}, and Self-RAG~\cite{Asai2023SelfRAGLT} enable the model to reason about the quality of the response and to coordinate multiple tool calls or retrieval steps to improve factual accuracy and robustness.

\paragraph{Tools and services.}
Practical deployment is as important as the underlying models. A production RAG assistant requires reliable hosting infrastructure, including vector stores and APIs, and may use specialised accelerator hardware (e.g.\ Groq inference engines~\cite{groq,park2025}) to reduce latency. Integration with existing collaboration tools is also important for adoption. Standards such as the Model Context Protocol (MCP)~\cite{mcp} provide a mechanism for language models to call external services or to be exposed as callable tools themselves, enabling interoperability across platforms. The infrastructure and integrations implemented in \ours are described in \cref{sec:core,sec:decisions}.

\subsection{Related Work}
Scientific institutions and research domains are beginning to adopt RAG for navigating complex, domain-specific literature. 

\paragraph{High Energy Physics.} The HEP community has begun to adopt LLMs augmented with retrieval, spanning documentation assistants, agentic analysis tools, and dedicated benchmarks. Among documentation-facing assistants, the AccGPT initiative at CERN is developing a knowledge-centric chatbot for interactive access to internal document collections, and supported the early development of \ours~\cite{rehm2025accgpt}. The CMS Archi system similarly combines the ingestion of heterogeneous scientific and institutional sources with configurable agents for retrieval and reasoning, and has been deployed to support CMS computing operations~\cite{Lugato:2026osi}. Related experiment-facing efforts include DUNE-GPT for the DUNE experiment~\cite{Rafique:2025xeo}, systems for digesting accelerator-facility electronic logbooks~\cite{Sulc:2024ssg}, and LHCb-focused retrieval systems that exploit article hierarchies and knowledge graphs to recover cross-document structure from particle-physics literature~\cite{McGreivy2025Forest}. More recently, MITRA provides access to internal analysis documentation at CMS~\cite{Mallampalli:2026hrl}.
Beyond documentation retrieval, a parallel line of work couples retrieval with agentic reasoning to assist with analysis itself. The Dr.~Sai project for BESIII combines the domain-adapted Xiwu language model with retrieval, knowledge graphs, and executable tools~\cite{SaiZZD2024,Xiwu}. MARVEL, from the LIGO gravitational-wave collaboration, is a multi-agent framework that pairs open LLMs with structured retrieval to navigate unstructured scientific data~\cite{MARVEL}. Recent systems also automate collider-analysis workflows end to end using LLM-powered agents~\cite{Gendreau-Distler:2025fsj,Plehn2026MadAgents,Moreno:2026mqk}. Large language models have begun to support physics theory itself, for example in Beyond-Standard-Model model building~\cite{Saad:2026pan}.
As these systems proliferate, dedicated benchmarks are appearing. Collider-Bench evaluates whether LLM agents can autonomously reproduce published LHC analyses from papers and open-source tools~\cite{colliderbench}, complementing the document-grounded evaluation developed for \ours (\cref{sec:evaluation}).
 
\paragraph{Beyond HEP.} The same pattern is emerging across the sciences and in industry. Retrieval-augmented generation, increasingly combined with LLM agents that plan and act over multiple steps~\cite{gao2025ragreasoning}, is being applied to domains from biomedicine~\cite{he2025biomedragsurvey} and chemistry~\cite{Zhong2025BenchmarkingRG} to other areas of physics, such as conversational agents for cosmology simulation codes~\cite{clapp2025}, and has moved from research prototype to production tool in enterprise settings, where organisations deploy it for citation-grounded assistance over proprietary corpora and major technology providers run it in production to improve factual grounding~\cite{wsj2024rag,wired2024rag,time2024patrick}. Adoption nonetheless remains uneven, with practitioners reporting persistent challenges in data preprocessing, security, and evaluation~\cite{brehme2025industry}.

%% file: sections/3.atlasdocs.tex
\section{ATLAS Information Ecosystem}
\label{sec:atlasdocs}
The challenge of documentation and information flow in ATLAS has been recognised since the early years of the experiment. The 2003 collaborative tools white paper already warned of the risks posed by fragmented communication channels and poorly integrated documentation~\cite{goldfarb2003atlas}. By 2010, a dedicated study, ``The use of the TWiki Web in ATLAS,'' reported more than 7,000 wiki pages authored by hundreds of contributors, and highlighted both the central role of the wiki and the difficulties of maintaining consistency and quality across such a large, collaboratively written system~\cite{amram2010twiki}. Fifteen years later the scale has multiplied, with more than 35,000 TWiki pages created alongside numerous additional repositories. The 2025 ``Big ATLAS Survey'' confirmed that the fundamental issues remain, with a clear finding being that \emph{``searchability is a common weakness of all tools''}~\cite{atlas2025survey}.

The problem is therefore not a lack of documentation or effort. On the contrary, ATLAS produces enormous volumes of material and invests significant resources in educating collaborators on how to use its software and computing systems. This includes centralised tutorials, self-guided documentation, and structured training programmes that have been organised regularly since 2004~\cite{atlas2025run3sc}. However, maintaining this documentation over time presents additional challenges. Information must be continually updated as software and workflows evolve, and undocumented expertise is regularly lost as collaborators leave the experiment or students complete their degrees. Together with the fragmentation of information across many systems, each with its own conventions and limited cross-linking, these factors remain a major barrier to the effective reuse and discovery of existing knowledge.

The following section summarises the major platforms currently used to capture and disseminate information within ATLAS. This overview focuses on documentation systems primarily used for off-detector work. In contrast, operations teams rely heavily on electronic logbooks and run-control interfaces, which are outside the scope of this discussion. 
These teams are already highly effective, as evidenced by ATLAS's strong data-taking efficiency and data-quality performance. 
For this reason, we do not target operations workflows in the present work, and would only consider extending \ours{} into these domains once the system has demonstrated consistently robust performance in other areas and reached the level of maturity required for mission-critical tasks.
Our emphasis is therefore on the fragmented documentation landscape that supports analysis and software development, where discoverability and unification remain significant challenges.

\noindent \textbf{TWiki.}  
TWiki~\cite{twiki} has historically been the main wiki for ATLAS, with over 35,000 pages of notes, instructions, and guides, and is internal to the collaboration. Content is unevenly maintained and written in an inconsistent style. A well-known issue is that much of the material is outdated or no longer accurate, as pages are seldom retired or revised once their authors move on. API access is not available, and the search mechanism is poor. It performs literal keyword matching, largely over page titles, with no relevance ranking, full-text indexing, or semantic matching, so locating the correct page typically requires already knowing its name. A custom back-end was developed by CERN IT for ATLAS, but the flat tree structure makes it difficult to distinguish between private or temporary pages and those maintained by groups. Despite these limitations, TWiki remains the single largest body of ATLAS documentation and is therefore included in \ours. The initial corpus includes pages created at any time, provided that they were modified within the preceding five years (\cref{sec:core}). This modification-date cutoff is a pragmatic, precision-oriented choice intended to limit false-positive retrievals from the much larger body of older, unmaintained material. Some older pages retain historical value, but identifying that subset reliably, for example through ownership and link-based signals, would require a dedicated corpus-curation study beyond the scope of the present work.

\noindent \textbf{Markdown Document (MkDocs) sites.}  
MkDocs~\cite{mkdocs} is a static site generator that builds documentation portals from \textit{Markdown} files stored in a source code management system.
They are lightweight documentation portals that are increasingly adopted as a replacement for TWiki. These sites are typically hosted in GitLab~\cite{gitlabcern}, which makes them easier to version, review, and maintain through standard software workflows, while remaining within the CERN ecosystem. As a more recent alternative to TWiki pages, approximately 10,000 pages of documentation are available.
While MkDocs provides a more modern and sustainable alternative, when \ours was first deployed in late 2023 the sites were heterogeneous and decentralised across the collaboration, with limited central indexing, and cross-site searching was not available.\footnote{A new ATLAS documentation search with cross-site functionality was subsequently developed and made available to the collaboration in May 2026.} MkDocs content is a mix of public and collaboration-internal material. Given its growing role and clean programmatic access, it is included in \ours, with documents collected via the GraphQL API~\cite{graphQL}.

\noindent \textbf{CERN Document Server (CDS).}  
The CERN Document Server (CDS)~\cite{cds} serves as the archive of internal notes, public papers, and conference reports. Internal notes are often supporting documents for public results and can be ten times longer than the published papers detailing all supporting work and variation tests. Known issues include reliance on inconsistent user-supplied metadata, search functionality that does not effectively cover full text, and incomplete cross-linking between superseded versions. Despite these weaknesses, CDS provides API access to nearly 40,000 internal ATLAS documents, 15,000 conference reports, and 1,500 public papers. CDS therefore spans both public material (journal papers and conference notes) and collaboration-internal notes. A limited subset of public ATLAS papers and notes has been imported into the \ours databases; the documents included in the present corpus are listed in \cref{sec:core}.

\noindent \textbf{Indico.}  
For meetings at all levels of the collaboration, Indico~\cite{indico} hosts agendas, contributions, slides, minutes, and recordings or links to recordings stored in CDS. Content ranges from very informal, such as collections of plots, to structured lectures. An API is available to access over 500,000 events, nearly 2 million contributions, and more than 1.5 million attachments (usually PDF files containing slide decks), stretching back to 1999. Material ranges from public to internal depending on the meeting. Given this very large volume, only a small, curated fraction is currently included in \ours. The present corpus uses the ATLAS continuing-education lectures (\cref{sec:core}), chosen for their pedagogical value and stable structure. Ingesting a broader selection of slides, for example from the main ATLAS weekly and physics meetings, is identified as a direction for future expansion in \cref{sec:conclusions}.

\noindent \textbf{CERN Groups and mailing lists.}
CERN's Group Management System provides group management and mailing-list functionality, replacing the earlier E-groups service model~\cite{cern_groups}. Mailing-list discussions remain a potentially valuable information source, but historical archives are incomplete and are not currently ingested into \ours. Much of the equivalent discussion is increasingly captured by ATLAS Talk.

\noindent \textbf{ATLAS Talk.}  
ATLAS Talk is a Discourse-based~\cite{discourse} forum platform for support, troubleshooting, and discussion across many topics. Categories can be linked to CERN mailing lists, allowing posts to be distributed by email and replies to be ingested back into ATLAS Talk, thereby bridging forum-style threaded discussion with existing mailing-list workflows. It provides improved search and archiving compared with email and is increasingly used as the central helpdesk for the collaboration, though adoption varies across groups. API access is available on request. Its content is collaboration-internal. Because each thread pairs a question with a community-validated answer, ATLAS Talk is a particularly valuable source and is included in \ours, where solved threads also serve as high-quality supervision for fine-tuning (\cref{sec:decisions}).

\noindent \textbf{Mattermost.}
Mattermost~\cite{mattermostcern} is a CERN-supported real-time messaging platform often used for technical Q\&A and coordination within groups. These conversations contain valuable collaboration knowledge, but their real-time, channel-based format makes them difficult to use as a stable documentation corpus. Self-hosting and support at CERN make it a preferred service to alternatives such as Microsoft Teams. Because of this, Mattermost is not ingested as a documentation source. Instead, \ours is integrated into Mattermost as a conversational interface (\cref{sec:decisions}).

\noindent \textbf{JIRA.}
JIRA~\cite{jira} is used as an issue-tracking system, for example to request sets of simulated events and to manage ongoing tasks in some groups. It enables structured workflows and accountability, but its use is not standardised across the collaboration and adoption varies widely. Owing to this inconsistent, group-dependent usage, JIRA is not currently included in the \ours corpus.

\noindent \textbf{Code repositories.}
CERN GitLab~\cite{gitlabcern} repositories (and some GitHub mirrors) host code, README files, and issues with embedded documentation. While co-location of code and documentation improves consistency, documentation quality and discoverability depend strongly on developer practices. 
In parallel, the LXR source browser~\cite{lxr} provides cross-referenced, web-based navigation of the Athena and AnalysisBase codebases, enabling full-text search and symbol-level exploration of C++ and Python implementations\footnote{\url{https://acode-browser.usatlas.bnl.gov/}}. Although widely used for debugging and software development, LXR is not integrated with other documentation systems and has no programmatic API, limiting automated retrieval or linking. Source code is therefore not directly indexed in the present version of \ours, which targets documentation rather than code. Code-aware assistance is discussed as future work (\cref{sec:conclusions}).

\noindent \textbf{Glance.}
Glance~\cite{glance} is a metadata and bookkeeping system for membership, appointments, and efforts to publish scientific results, ensuring central records are maintained. Its structured records on roles, appointments, and publication workflows are an attractive source of current collaboration knowledge. Although Glance is not currently ingested into \ours, structured API access has been available since October 2026, making it a near-term candidate once its records are converted into retrievable natural-language descriptions.

Together these platforms provide comprehensive coverage of ATLAS activities, but in practice they form a highly fragmented ecosystem. Documentation is written in many voices, with little editorial oversight, and crucial information is dispersed across platforms that are difficult to search in combination. The result is a paradox of abundance: the necessary information exists, but locating and integrating it into daily work remains a significant bottleneck that reduces efficiency and increases reliance on informal communication. \ours{} is designed to address this gap by providing a unifying conversational interface across existing systems, complementing rather than replacing the substantial documentation and training efforts already in place.


%% file: sections/4.benchmark.tex
\section{Benchmarking and Evaluation of RAG}
\label{sec:evaluation}

Evaluating retrieval-augmented generation (RAG) systems in the context of ATLAS requires a domain-specific benchmark. General-purpose evaluation frameworks such as RagBench~\cite{ragBench} are valuable for open-domain natural language processing, but they do not capture the specialised terminology, document structures, and interdisciplinary content characteristic of large-scale scientific collaborations. To address this gap, \texttt{chATLAS\_Benchmark}~\cite{chatlas_benchmark}, a purpose-built evaluation framework tailored to ATLAS documentation, was developed to provide a reproducible basis for comparison.

\subsection{Benchmark Framework}

The benchmark follows a document-grounded synthetic-question generation paradigm, similar in spirit to recent RAG evaluation frameworks such as RAGAS~\cite{RAGAS} and ARES~\cite{ARES}. Questions are generated directly from the target document corpus rather than curated externally, ensuring close alignment between the evaluation data and the deployed knowledge base. The evaluation is deliberately restricted to non-adversarial scenarios intended to resemble typical ATLAS information requests.

Each source document or document chunk is supplied to an LLM together with a persona-conditioned prompt instructing it to pose a realistic question that the passage answers and to compose a reference answer grounded only in the supplied text. The originating document is recorded as the ground-truth source for that pair. Multiple personas simulate different user profiles, including early-career students, experienced analysts, and domain experts. The exact prompts and implementation are distributed with the \texttt{chATLAS\_Benchmark} package~\cite{chatlas_benchmark}. Candidate pairs are automatically checked for relevance, answerability from the source, and answer correctness, with representative samples inspected manually.

\subsection{Training and Evaluation Datasets}
\label{sec:retrievalbenchmark}

Two kinds of question--answer (QA) data are used and should be distinguished. For TWiki and Git-based documentation, QA pairs are generated synthetically with the framework above. For ATLAS Talk, real user questions and their corresponding community answers provide higher-quality semantic supervision. Only threads marked as solved or answered are retained, and an additional \texttt{gpt-5-mini} filtering step removes incomplete or low-quality examples; this filtering was checked by manual inspection of representative samples.

The controlled embedding study uses three documentation sources: TWiki, Git-based documentation, and ATLAS Talk. For each source, 4,000 QA pairs are used for fine-tuning, 400 for validation, and 1,000 for testing. The source-specific models therefore use equally sized training sets, preventing corpus size from becoming a confounding variable. The combined model uses all 12,000 training pairs. The 4,000-pair choice was set by the smallest available source and is supported by the scaling study in \cref{app:details_benchmark}: the largest gains occur in the first few thousand pairs and diminish beyond approximately 10,000 pairs. A model without fine-tuning provides the baseline.

The questions, and hence the QA records, are mutually exclusive across the training, validation, and test partitions.

Synthetic, document-grounded questions provide an unambiguous retrieval target and make controlled comparisons practical, but they do not fully reproduce the distribution, ambiguity, or multi-document nature of real user queries. The ATLAS Talk set partly mitigates this limitation because its questions originated with users, while production-log examples in \cref{app:examples} provide a separate qualitative view of deployed usage.

\subsection{Evaluation Metrics}

Retrieval and generation performance are evaluated with complementary metrics. The \emph{document match} metric checks whether the labelled source document appears in the retrieved set; top-$k$ accuracy reports the fraction of questions for which it appears among the first $k$ results. The weighted document-match score additionally rewards correct sources placed nearer the top of the ranking. Lexical metrics, including exact match, F1, and ROUGE~\cite{ROUGE}, assess surface overlap between predicted and reference answers, while embedding-based semantic similarity measures alignment in meaning beyond exact wording. These retrieval and text-based indicators are independent of LLM judgement.

For the generator comparison in \cref{sec:decisions}, the framework additionally uses an \emph{LLM-as-judge} correctness metric. An independent model receives the question, reference answer, and generated answer and returns a binary judgement of whether the generated answer correctly addresses the question. This metric captures answer quality that surface overlap may miss, but it also inherits the limitations of the chosen judge; the judge model and its settings are therefore reported with the comparison.

This evaluation framework guides the design decisions presented in \cref{sec:decisions}.

%% file: sections/5.core.tex
\section{Core Infrastructure}
\label{sec:core}

The current implementation of \ours is built around a RAG architecture, with many components deployed locally at CERN. The system is designed to be modular, allowing different components (retrievers, embedding models, generators) to be interchanged as development progresses. The software is released as a set of open-source Python packages covering data scraping, embedding and vector-store management, and retrieval chains~\cite{chatlas_scrape,chatlas_embed,chatlas_chains}, alongside the benchmarking framework of \cref{sec:evaluation}~\cite{chatlas_benchmark}.

The backend consists of a PostgreSQL~\cite{rowe1987postgres} database with the \texttt{pgvector}~\cite{pgvector} extension to support efficient similarity search. PostgreSQL is provided as part of CERN's Platform-as-a-Service (PaaS)~\cite{paas} offering, which provides institutional monitoring, support, and scalable infrastructure.

The Flask frontend authenticates users through CERN Single Sign-On using OpenID Connect and CERN's Keycloak discovery endpoint. After the OAuth flow, the returned identity and authorisation information, including the user's permissions, is held in the application session and used to verify ATLAS membership before access to the search and assistant routes is granted. A separate local-development mode bypasses OAuth and uses a test logging database.

The document corpus, detailed in \cref{sec:atlasdocs}, includes a subset of ATLAS information sources. The counts below and in \cref{tab:retrieval_stats} are a single operational snapshot from late 2025:
\begin{itemize}
    \item TWiki: Pages modified within the last 5 years, totalling 4,450.
    \item MkDocs: Markdown pages modified within the last 2 years, totalling 9,982.
    \item CDS: 1,914 ATLAS papers, 1,250 ATLAS CONF notes, and 39 ATLAS PUB notes.
    \item Indico: 217 ATLAS education lectures organised as part of the continuing education programme.
    \item ATLAS Talk: 13,824 question--answer threads.
\end{itemize}

Preprocessing steps normalise formatting, strip boilerplate, and attach metadata such as authorship, date, description, and working group. A parent-child structure is used with parent-chunks of 1024 tokens divided into four 256-token children, as detailed in \cref{tab:retrieval_stats}. The per-source counts listed above are numbers of source documents, whereas the table reports the parent and child chunk counts obtained after segmentation. During generation, the full parent document is passed to the LLM for every child document found by the vector search.
The TWiki and MkDocs content is updated weekly via a scheduled GitLab CI pipeline. 

Both the document chunks stored in the database and the user queries are mapped into a vector space using a domain-adapted embedding model. The base encoder is \texttt{sentence-transformers/all-mpnet-base-v2}~\cite{reimers2019sentencebert,song2020mpnet}, which produces a 768-dimensional representation as part of its pretrained architecture. The dimensionality is inherited from the chosen model. To improve coverage of collaboration-specific terminology, acronyms, and jargon (including everyday words whose ATLAS meaning is highly specialised, such as ``track'', ``cluster'', ``fit'', ``tag'', or ``trigger''), the model was further fine-tuned using question--answer pairs from all sources. The fine-tuned model is used consistently for encoding both the database entries and incoming queries.

Before retrieval, each query first passes through an intent-routing and safety layer (\cref{sec:routing}) that screens out unsafe or out-of-scope requests and directs person-lookup queries to the appropriate resource. In assistant mode, the query is then normalised by a lightweight LLM step that corrects spelling and resolves conversational references into a self-contained query before it is embedded and searched (\cref{sec:decisions}).

For retrieval, a hybrid strategy is employed. Dense retrieval is performed using vector similarity search over embeddings stored in \texttt{pgvector}, with cosine similarity as the distance metric.
In parallel, sparse retrieval is implemented using PostgreSQL's built-in full-text search (\texttt{tsvector}/\texttt{tsquery}), which applies a TF-IDF-based ranking~\cite{sparckjones1972} similar in form to BM25 while providing tighter integration with the database back-end.
A configurable parameter defines the number of matches returned per source. The defaults are shown in \cref{tab:retrieval_stats}. Results from the two retrievers are merged using Reciprocal Rank Fusion (RRF, \cref{eq:rrf}) with $k = 60$.

\begin{table}[ht]
\centering
\caption{Operational snapshot of dataset segmentation and retrieval statistics for the major ATLAS documentation sources from late 2025. The production corpus is refreshed weekly. Parent and child chunks refer to the hierarchical segmentation used. $N_{dense}$ and $N_{sparse}$ refer to the number of results returned per source from the dense and sparse retrieval, respectively. Both are user-adjustable, with the deployed defaults shown.}
\label{tab:retrieval_stats}
\begin{tabular}{lcccc}
\toprule
Source & Parent Chunks & Child Chunks & $N_{dense}$ & $N_{sparse}$ \\
\midrule
TWiki        & 148,707 &   661,153 & 4 & 3 \\
MkDocs       &  45,838 &   201,681 & 4 & 3 \\
CDS          & 224,965 & 1,029,826 & 4 & 3 \\
ATLAS Talk   &  29,294 &   233,162 & 4 & 3 \\
Indico       &   6,491 &    32,969 & 4 & 3 \\
\bottomrule
\end{tabular}
\end{table}

The generative component is provider-agnostic, with configured support for open-weight models served through the Groq API and hosted OpenAI models. The system is intended to operate with locally deployable open-source models hosted at CERN in future releases. The specific models used for each task are summarised in \cref{sec:decisions}.

\begin{figure}[htbp]
\centering
\includegraphics[width=\textwidth]{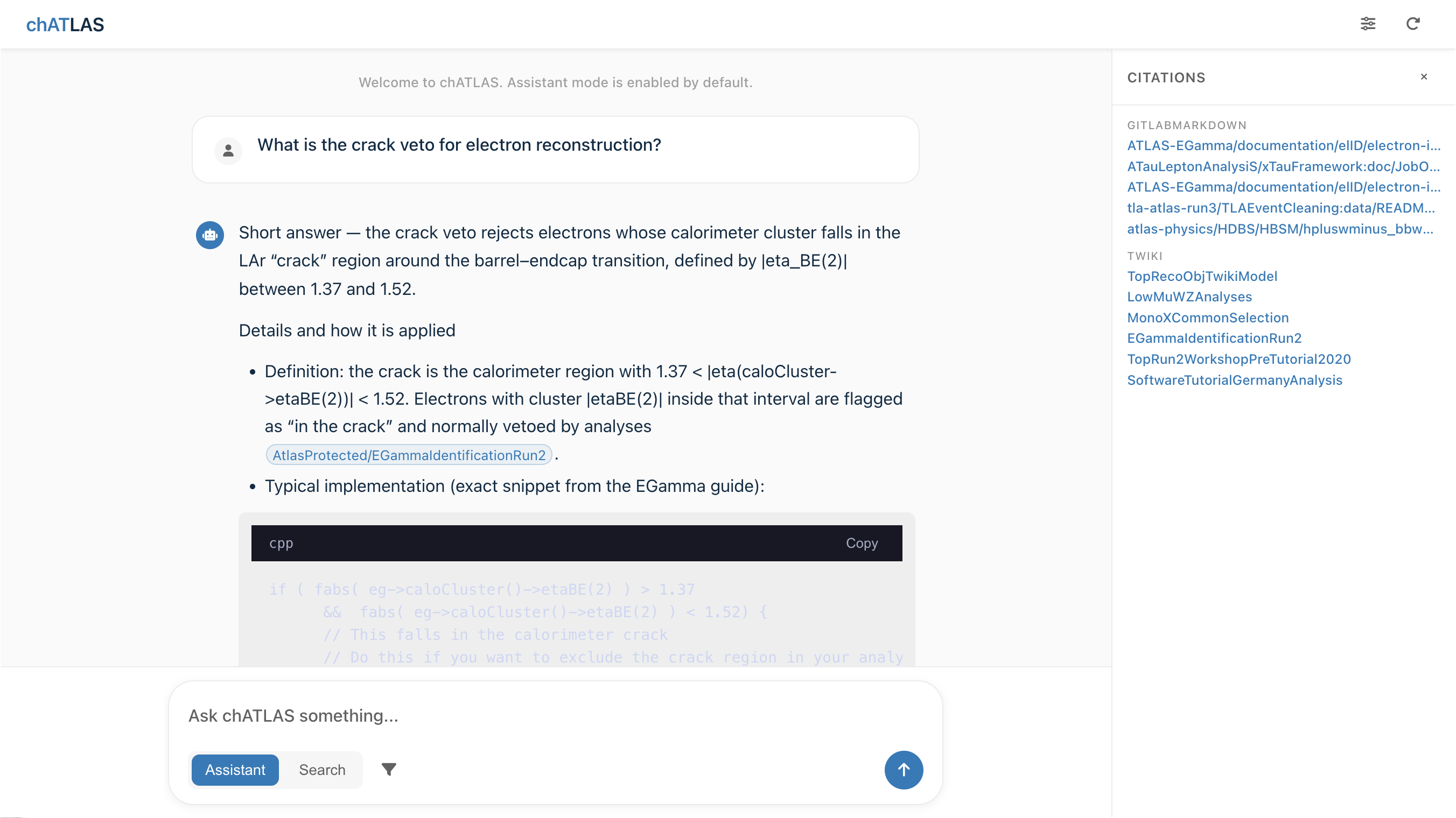}
\caption{The \ours web interface in assistant mode, responding to the query ``What is the crack veto for electron reconstruction?''. The generated answer includes inline source attribution and a code snippet drawn from the cited documentation. The right-hand panel lists the retrieved sources grouped by documentation corpus (GitLab Markdown and TWiki in this example), each linking to the original document. The control at the bottom switches between assistant and search modes.}
\label{fig:ui_example}
\end{figure}

The user interface is implemented using a lightweight Flask web app with custom JavaScript components to provide a conversational user interface.\footnote{\ours is available to ATLAS collaboration members at \url{https://chatlas-flask-chatlas.app.cern.ch} (CERN authentication required).} This interface supports both search mode and assistant mode. In search mode, the system performs retrieval only and does not invoke the generative LLM. Only the text embedding model and the sparse retriever are used, and the top ranked sources given the query are listed with hyperlinks to the location of the original document along with the corresponding similarity score. In assistant mode, the top sources are passed to an LLM for chat-style dialogue where a subset of sources is displayed with each generated answer. These assistant-mode interactions are \emph{multi-turn}. \ours retains the recent conversation history (the last several query--response pairs) and uses it to interpret follow-up questions, which are rewritten into self-contained queries before retrieval (\cref{sec:decisions}). The user interface also provides checkboxes for selecting source corpora and categories, controls for filtering by last-modified date, and parameters determining the number of results retrieved from each corpus. Each \ours response is accompanied by a binary feedback mechanism (thumbs-up and thumbs-down) that is sent along with the query thread to a feedback database. A representative selection of real user queries and the responses they received, drawn from the production query log, is presented in \cref{app:examples}. \Cref{fig:ui_example} shows the interface in assistant mode.

Connection with collaboration tools such as Mattermost has been successfully achieved, enabling \ours to be called directly within a chat.

Launched in November 2023 and in continuous production use on CERN's PaaS infrastructure since April 2024, \ours has now operated for over two years, with sustained engagement across the collaboration. When development began, CERN-hosted LLM services were not yet available, and a central design goal was therefore to deliver useful multi-source search at minimal operating cost. Usage is split between the assistant (chat) and search modes, and the introduction of the Mattermost integration in mid-2025 produced a noticeable increase in engagement. At the time of its launch, \ours was, to our knowledge, the only ATLAS service supporting simultaneous search across multiple internal documentation sources, including TWiki and GitLab-hosted MkDocs sites.

%% file: sections/6.design.tex
\section{Design Decisions and Insights}
\label{sec:decisions}

The development of \ours required a series of design choices that were guided by the characteristics of the ATLAS documentation ecosystem and validated through dedicated benchmarks (\cref{sec:evaluation}). In this section we summarise the main decisions, their motivations, and the outcomes of experimental studies.

\subsection{Source Synchronisation and Hierarchical Chunking}
The document corpus (\cref{sec:atlasdocs}) was pre-processed into retrievable units with a parent-child structure. Preliminary evaluations indicated that this hierarchical organisation outperformed a flat representation, improving retrieval consistency without adding complexity. Varying the child chunk size affected document match rates by approximately 5\%, while the inclusion of additional metadata, such as the Git repository description for MkDocs pages, increased the rate of correct document retrievals by about 15\%.

Having programmatic API access to documentation sources is essential to enable reliable updates and automated synchronisation. Initially, TWiki content was retrieved using \texttt{wget}, which proved brittle and inefficient for large-scale, regular updates. The current pipeline instead uses the CERN-maintained filesystem-backed TWiki source available to ATLAS, enabling automated weekly indexing without crawling the live web interface.

\subsection{Fine-Tuned Text Embeddings}

For retrieval, both document chunks and user queries are mapped into a common semantic vector space using sentence embeddings. The current embedding model is \texttt{sentence-transformers/all-mpnet-base-v2}, selected based on retrieval performance using the benchmark described in \cref{sec:retrievalbenchmark}. We additionally investigated more recent state-of-the-art embedding models, including \texttt{google/embeddinggemma-300m}, as well as models externally adapted to physics literature such as PhysBERT~\cite{PhysBERT}. We found that neither larger general-purpose embedding models nor models specialised for public physics papers consistently outperformed the domain-adapted embedding model on ATLAS internal documentation.

To further adapt the embedding model to the ATLAS domain, contrastive fine-tuning was performed using the training datasets defined in \cref{sec:retrievalbenchmark}. These combine real, expert-validated ATLAS Talk question--answer pairs with synthetic pairs generated from the remaining documentation sources. This supervision teaches the model to place semantically related questions and answers close together in the embedding space.
\begin{table}[ht]
\centering
\caption{Training configuration for the embedding model fine-tuning.}
\label{tab:finetune_config}
\begin{tabular}{lc}
\toprule
\textbf{Parameter} & \textbf{Value} \\
\midrule
Base model & \texttt{sentence-transformers/all-mpnet-base-v2} \\
Training epochs & 3 \\
Optimizer & PyTorch AdamW \\
Learning rate & $5 \times 10^{-7}$ \\
Train batch size (per device) & 16 \\
Eval batch size (per device) & 16 \\
Gradient-accumulation steps & 4 \\
Effective optimisation batch size & 64 \\
In-batch negative pool & 16 \\
Warmup ratio & 0.1 \\
Random seed & 42 \\
Precision & 16-bit (FP16) \\
Batch sampler & \texttt{BatchSamplers.NO\_DUPLICATES} \\
Evaluation steps & 100 \\
Similarity function & Dot product \\
Scaling factor & 1.0 \\
GPU hardware & NVIDIA A100 (40\,GB) \\
\bottomrule
\end{tabular}
\end{table}

Fine-tuning employed the \texttt{CachedMultipleNegativesRankingLoss}~\cite{reimers2019sentencebert}, a variant of the Multiple Negatives Ranking loss~\cite{Henderson2017EfficientNL} that maximises similarity between positive question--answer pairs while treating the remaining answers within each batch as negative examples. We trained for three epochs using the PyTorch AdamW optimiser, a learning rate of $5 \times 10^{-7}$, a warmup ratio of 0.1, and a random seed of 42. Training used a per-device batch size of 16 with four gradient-accumulation steps, corresponding to an effective optimisation batch size of 64 on a single GPU; the in-batch negative pool contained 16 examples per forward pass. Training was performed using the Hugging Face \texttt{Trainer}. The complete training configuration is summarised in \cref{tab:finetune_config}.

\begin{figure}[htbp]
    \centering
    \includegraphics[width=0.75\linewidth]{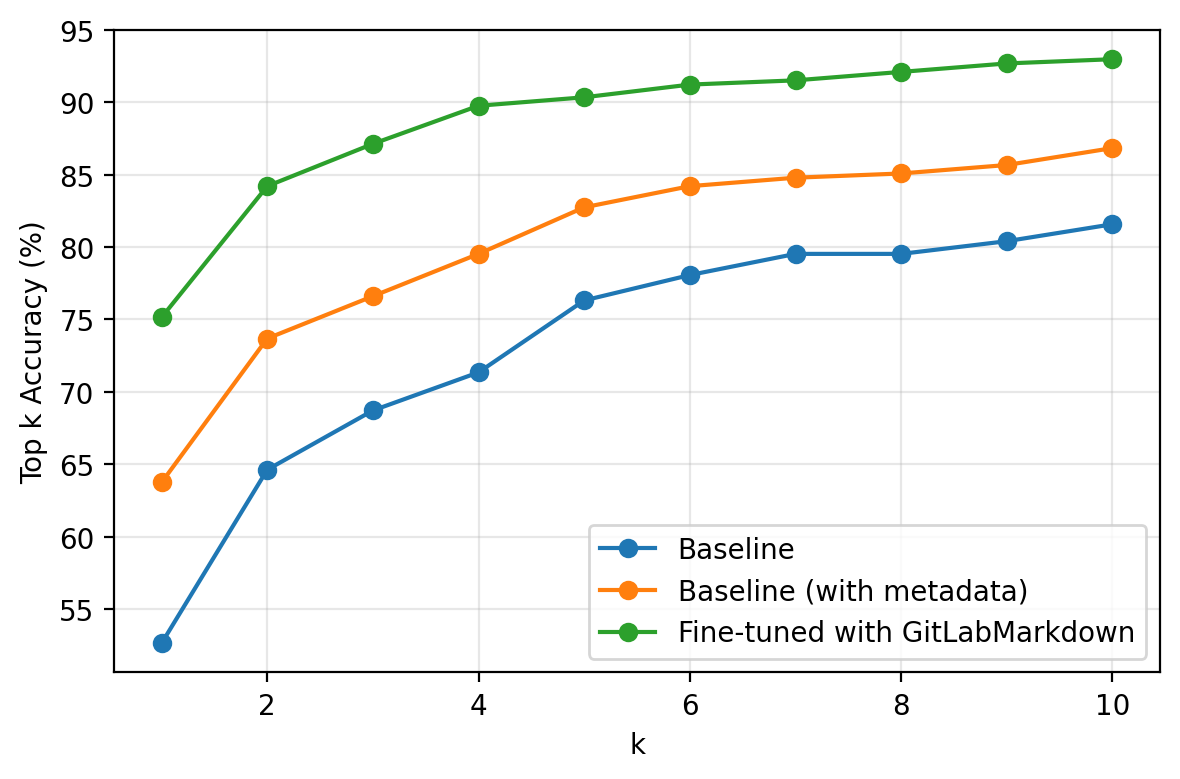}
    \caption{Top-$k$ retrieval accuracy illustrating successive improvements to the embedding pipeline. Adding document metadata substantially improves retrieval performance relative to the baseline embedding model, while domain-specific fine-tuning provides a further improvement. For illustration, the fine-tuned model was trained and evaluated on the GitLab Markdown benchmark.}
    \label{fig:embeddingmodel}
\end{figure}

The embedding model evolved through several iterations as the retrieval benchmark matured. Early deployments used \texttt{all-MiniLM-L6-v2}, followed by \texttt{multi-qa-mpnet-base-dot-v1}. Domain-specific fine-tuning, metadata augmentation, and subsequent evaluation using the proposed retrieval benchmark led to the current embedding model. The contributions of these successive improvements are summarised in \cref{fig:embeddingmodel}.

\subsection{Cross-Source Retrieval Evaluation}
\label{sec:crosssourceeval}
\begin{figure}[htbp]
    \centering

    \begin{subfigure}{0.48\linewidth}
        \centering
        \includegraphics[width=\linewidth]{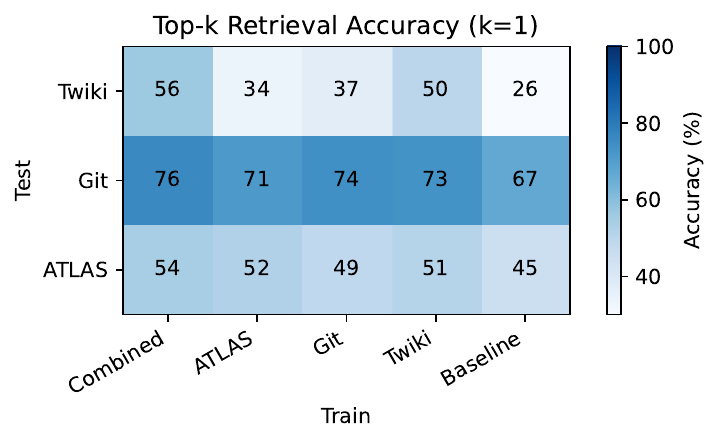}
        \caption{Cross-source retrieval accuracy at $k=1$.}
        \label{fig:embedding_crosssource_k1}
    \end{subfigure}
    \hfill
    \begin{subfigure}{0.48\linewidth}
        \centering
        \includegraphics[width=\linewidth]{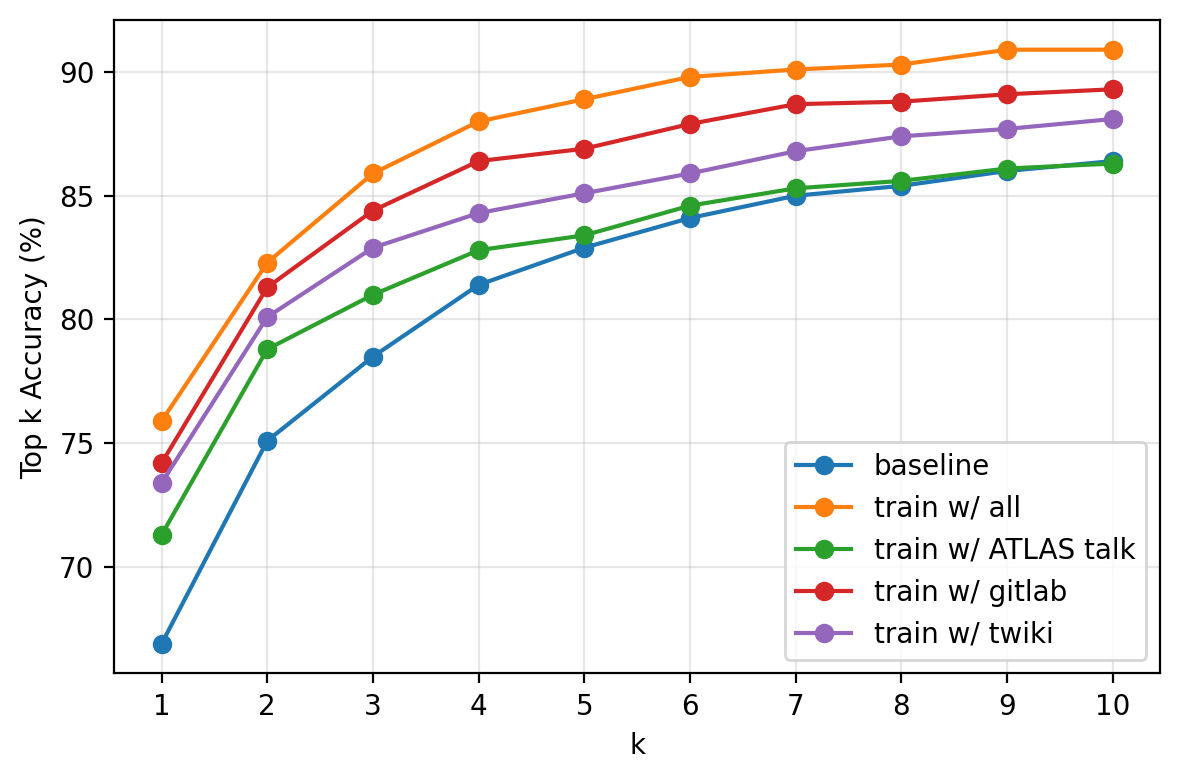}
        \caption{Top-$k$ retrieval accuracy as a function of $k$.}
        \label{fig:embedding_crosssource_k}
    \end{subfigure}

    \caption{Cross-source evaluation of embedding models before applying contextual retrieval. (a) Top-$1$ retrieval accuracy for each combination of evaluation (`test') source (rows) and fine-tuning (`train') source (columns). Source-specific models were fine-tuned on the corresponding documentation source, while the combined model was fine-tuned on data from all three documentation sources. Baseline denotes the untuned embedding model. (b) Top-$k$ retrieval accuracy as a function of retrieval depth for the GitLab evaluation set, corresponding to the second row of (a). The combined model consistently achieves the highest retrieval accuracy across all values of $k$, while source-specific fine-tuning provides intermediate improvements over the untuned baseline. Retrieval accuracy increases rapidly for the first few retrieved documents, with the largest improvement occurring from $k=1$ to $k=2$, before exhibiting diminishing returns beyond approximately $k=5$.}
    \label{fig:embedding_crosssource}
\end{figure}

\Cref{fig:embedding_crosssource} summarises the top-$k$ retrieval accuracy ($k=10$) for all train--test combinations. Fine-tuning consistently improves retrieval performance relative to the baseline, even when evaluated on documentation sources different from those used for training. This cross-source transfer suggests that the embedding model learns domain-specific retrieval patterns and ATLAS terminology that generalise across heterogeneous documentation sources. 

Models fine-tuned on a single source exhibit the strongest performance when evaluated on that same source, while retaining a substantial fraction of their gains when applied to other sources. The combined training configuration yields the most stable and consistently strongest performance across all evaluation sets, indicating that exposure to multiple documentation styles and vocabularies produces a more robust embedding model for production deployment.

Apart from the combined training configuration, the best performance is generally observed when models are evaluated on the same source used for fine-tuning. While this partly reflects source-specific vocabulary and document structure, differences in corpus size and technical complexity also play a significant role. In particular, the TWiki documentation corpus contains roughly an order of magnitude more retrievable documents than MkDocs or ATLAS Talk and spans a broader range of highly technical detector, software, and physics documentation, including long procedural pages and tabulated information. The larger and more technically diverse corpus therefore presents a substantially more challenging retrieval task than the smaller documentation sources. Nevertheless, fine-tuning consistently improves retrieval performance across all document sources.

\subsection{Contextual Retrieval}

The cross-source evaluation identified TWiki as the most challenging documentation source, motivating additional retrieval strategies beyond embedding fine-tuning. Unlike Git-based documentation or ATLAS Talk, TWiki pages often contain long procedural documentation, extensive tables, and highly specialised detector and software terminology, making individual document chunks difficult to interpret in isolation. We therefore hypothesised that enriching each chunk with document-level context would improve retrieval performance.

To address this, Anthropic's contextual retrieval technique~\cite{AnthropicContextualRetrieval} was implemented. For each document chunk, an LLM (\texttt{gpt-5-mini}) was provided with both the chunk and its parent document and tasked with generating a concise contextual description. The generated context was then embedded together with the original chunk as metadata, enriching the embedding space without modifying the embedding model itself. The implementation is part of the open-source \ours software packages~\cite{chatlas_embed,chatlas_chains}.
\begin{figure}
    \centering
    \begin{subfigure}[t]{0.48\textwidth}
        \centering
\includegraphics[width=\textwidth]{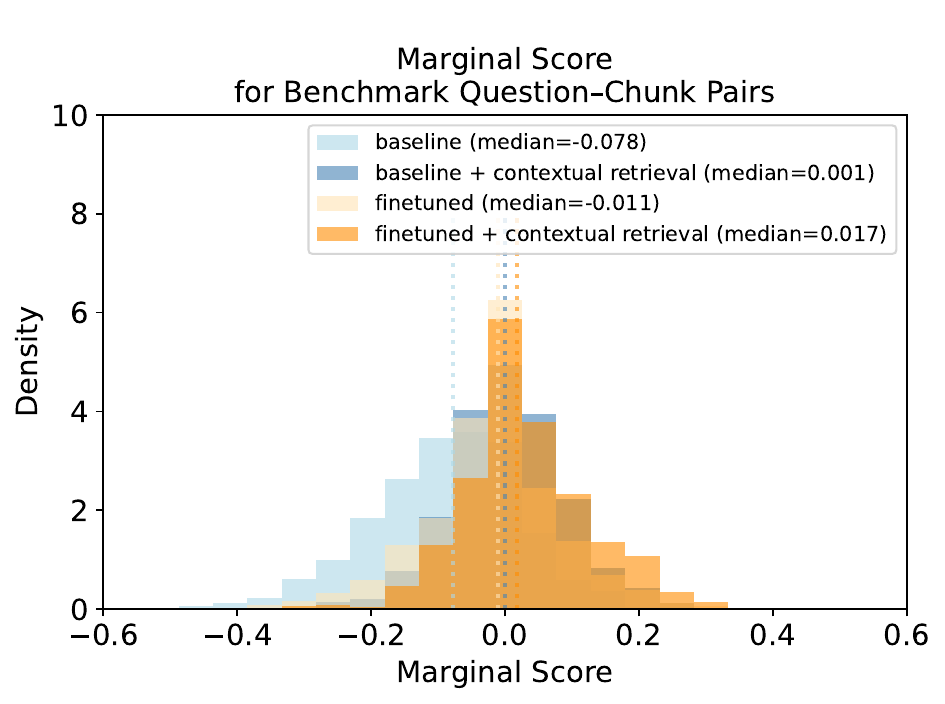}
        \caption{Marginal score}
        \label{fig:marginal_score}
    \end{subfigure}
    \hfill
    \begin{subfigure}[t]{0.48\textwidth}
        \centering
\includegraphics[width=\textwidth]
{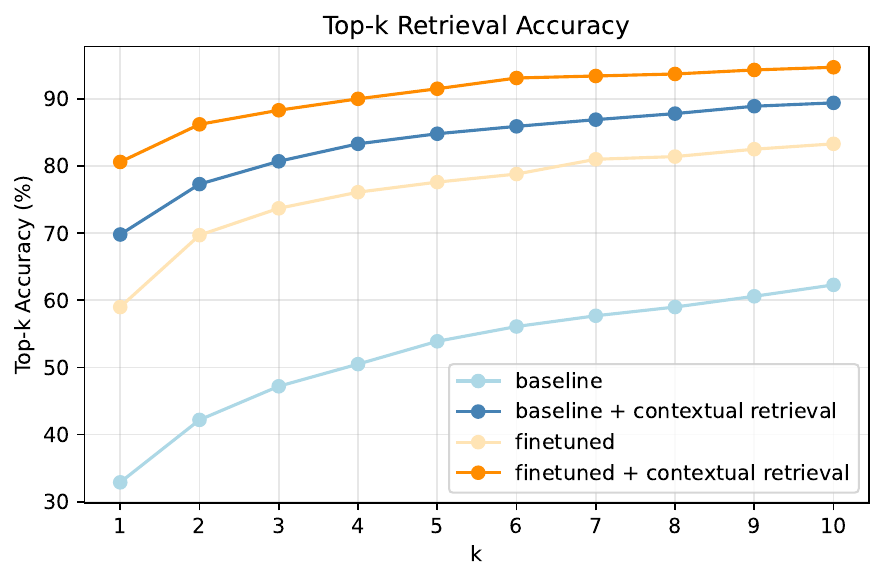}
        \caption{Top-$k$ retrieval accuracy for $k$ = 1 to 10}
        \label{fig:top_k_contextual_retrieval}
    \end{subfigure}

    \caption{Comparison of baseline and fine-tuned embedding models with and without contextual retrieval on the TWiki benchmark dataset. The left panel shows the marginal score distribution, defined as the difference between the highest cosine similarity score from the correct parent document and the strongest competing incorrect parent document. The right panel shows the top-$k$ retrieval accuracy for $k=1$ to $10$. Both metrics show that combining fine-tuning with contextual retrieval yields the best retrieval performance, improving top-1 accuracy by more than 140\% relative to the baseline model.
}

    \label{fig:contextual_retrieval}
\end{figure}

To better understand the retrieval improvements, we define a marginal score measuring the difference between the highest cosine similarity score of the correct parent document and that of the strongest competing incorrect parent document. The complete definition is given in \cref{appendix:contextual_retrieval}.

\Cref{fig:marginal_score} shows the distribution of the marginal score, which measures how much more strongly the retriever prefers the correct parent document over the strongest competing incorrect document. Contextual retrieval substantially shifts the marginal score towards larger values for both the baseline and fine-tuned embedding models, indicating improved separation between relevant and competing documents. The largest improvement is obtained by combining contextual retrieval with embedding fine-tuning.

\Cref{fig:top_k_contextual_retrieval} shows the corresponding top-$k$ retrieval accuracy. Consistent with the marginal score distributions, the fine-tuned model with contextual retrieval achieves the highest retrieval accuracy across all values of $k$. Interestingly, the baseline embedding model augmented with contextual retrieval outperforms the fine-tuned embedding model without contextual retrieval, demonstrating that document-level contextual information provides a larger performance gain than embedding fine-tuning alone.

The ordering of the top-$k$ retrieval accuracies closely matches the ordering of the marginal score distributions, suggesting that retrieval performance is primarily driven by increasing the separation between the correct parent document and competing alternatives. Together, these results support the hypothesis that contextual retrieval effectively bridges the semantic gap between user queries and highly technical scientific documentation, substantially improving retrieval on the challenging TWiki benchmark and reducing the performance gap observed relative to the other documentation sources in \cref{fig:embedding_crosssource}. The complementary improvements obtained by combining contextual retrieval with embedding fine-tuning further suggest that the two approaches capture different aspects of the retrieval problem.

\subsection{Vector Index Selection}

Dense retrieval is implemented using \texttt{pgvector}, with cosine similarity used to measure relevance between query and document embeddings. While cosine similarity defines the notion of proximity in the embedding space, practical retrieval over large corpora relies on nearest-neighbour search algorithms to efficiently identify candidate vectors. \texttt{pgvector} supports both exact nearest-neighbour search, which guarantees optimal results at high computational cost, and approximate nearest-neighbour (ANN) methods, including Hierarchical Navigable Small World (HNSW)~\cite{Malkov2016EfficientAR} and Inverted File with Flat Compression (IVFFlat)~\cite{IFVFlat}, which trade a small loss in recall for substantial speed gains.

After tuning the index and search parameters (\cref{tab:pgvector_searchparams}), HNSW and IVFFlat were found to deliver comparable retrieval accuracy and latency on the TWiki corpus, as well as similar index sizes. Dense retrieval using IVFFlat was therefore selected as the default. Grid searches over the number of dense and sparse candidates indicated that retrieving approximately $K=20$ dense candidates and $K_{\mathrm{text}}=10$ sparse candidates in total provided near-optimal performance. In the deployed system these totals are distributed across the selected documentation sources, with $N_{dense}=4$ dense and $N_{sparse}=3$ sparse candidates retrieved per source by default (\cref{tab:retrieval_stats}).

\begin{table}[htb]
\centering
\caption{Parameters chosen after tuning the \texttt{pgvector} search algorithms.
For HNSW, $m$ and $ef\_construction$ specify the maximum number of connections per layer and the size of the dynamic candidate list, respectively, used for constructing the multilayer graph for its index. When searching the HNSW index, $ef\_search$ specifies the size of the dynamic candidate list to use. The IVFFlat index divides vectors into $list$ lists when the index is built, and searches $probes$ of the lists closest to the search vector. $n_{entries}$ is the total number of embeddings in the database (only child chunks are embedded). In contrast to HNSW, the IVFFlat index needs to be rebuilt after adding documents. For comparison, \texttt{pgvector} default values are also shown.}
\label{tab:pgvector_searchparams}
\begin{tabular}{llcc}
\toprule
\textbf{Algorithm} & \textbf{Parameter} & \textbf{Value} & \textbf{\texttt{pgvector} default} \\
\midrule
HNSW & $m$                 &   16 &  16 \\
     & $ef\_construction$  &   64 &  64 \\
     & $ef\_search$        &  100 &  40 \\
\midrule
IVFFlat & $lists$   &  $1.05 \cdot \sqrt{n_{entries}}$   & - \\
        & $probes$  &  $\max(32, \, 0.14\cdot lists)$    & 1 \\
\bottomrule
\end{tabular}
\end{table}

\subsection{Hybrid Retrieval and Rank Fusion}
In the initial implementation, ranks from the dense (vector) and sparse (BM25) retrievers were combined by simple concatenation, ordering documents by their individual similarity or relevance scores. This baseline achieved moderate accuracy but suffered from inconsistencies when the two retrievers disagreed. Applying RRF (\cref{sec:rag}) resolved this by merging rankings according to the reciprocal of each document’s position across retrievers. Retrieval quality for this comparison is quantified by the \emph{weighted document-match score} of the \texttt{chATLAS\_Benchmark} framework~\cite{chatlas_benchmark}. For each benchmark query, a correct source document retrieved at rank $r$ among $N$ candidates scores $1-(r-1)/N$ (zero if it is not retrieved), so that correct documents ranked earlier score more highly, and the final score is the average over all queries. RRF improved this score from approximately 50\% to 65\%. A further stage using a neural cross-encoder re-ranker (Cohere Rerank 3.5~\cite{cohere2024rerank}) increased the same metric to around 71\%. Unlike the bi-encoder used for dense retrieval, which embeds the query and each document independently and compares the resulting vectors, a cross-encoder jointly encodes the query together with a candidate document in a single transformer pass. This allows token-level attention between the query and the document and therefore a more accurate relevance score, at the cost of one model evaluation per candidate document rather than a single cheap vector comparison. Despite the additional gain, the computational overhead of the neural re-ranker, requiring one forward pass per candidate document, made it impractical for large-scale or interactive use. For \ours, RRF was therefore adopted as the default approach, offering an effective balance between retrieval accuracy, latency, and implementation simplicity.

\subsection{Query Preparation and Self-Correction}

Generation is performed by a large language model conditioned on the retrieved passages. To improve robustness to noisy and conversational queries, the deployed assistant first passes each query through a lightweight LLM (currently OpenAI's \texttt{gpt-5-nano}) that corrects spelling and resolves conversational references (such as pronouns and follow-up questions) into a self-contained query. This rewritten query is the one that is embedded and searched. In parallel, the lexical (keyword) retriever applies character normalisation and stop-word removal when constructing its full-text query. These conservative query-handling steps are part of the running system.

More aggressive query- and answer-side strategies were studied but are not deployed. For query reformulation, Hypothetical Document Expansion (HyDE)~\cite{hyde2022} was explored. It improved recall, but frequently introduced hallucinated context in this domain, so the conservative strategy above was retained. A more aggressive LLM query-rewriting option is also implemented but is disabled by default. For answer quality, a self-evaluation layer was tested, in which an independent LLM judges candidate answers and triggers regeneration when quality falls below a threshold. Heterogeneous evaluation, using a judge model different from the generator, proved most effective, correcting over 15\% of initially incorrect responses in benchmark studies. These agentic loops add latency and complexity, so the deployed system uses a single-pass pipeline, with the self-correction machinery retained for possible future use.

\subsection{Generator Selection}
\begin{figure}[htbp]
\centering
    \includegraphics[width=0.78\linewidth]{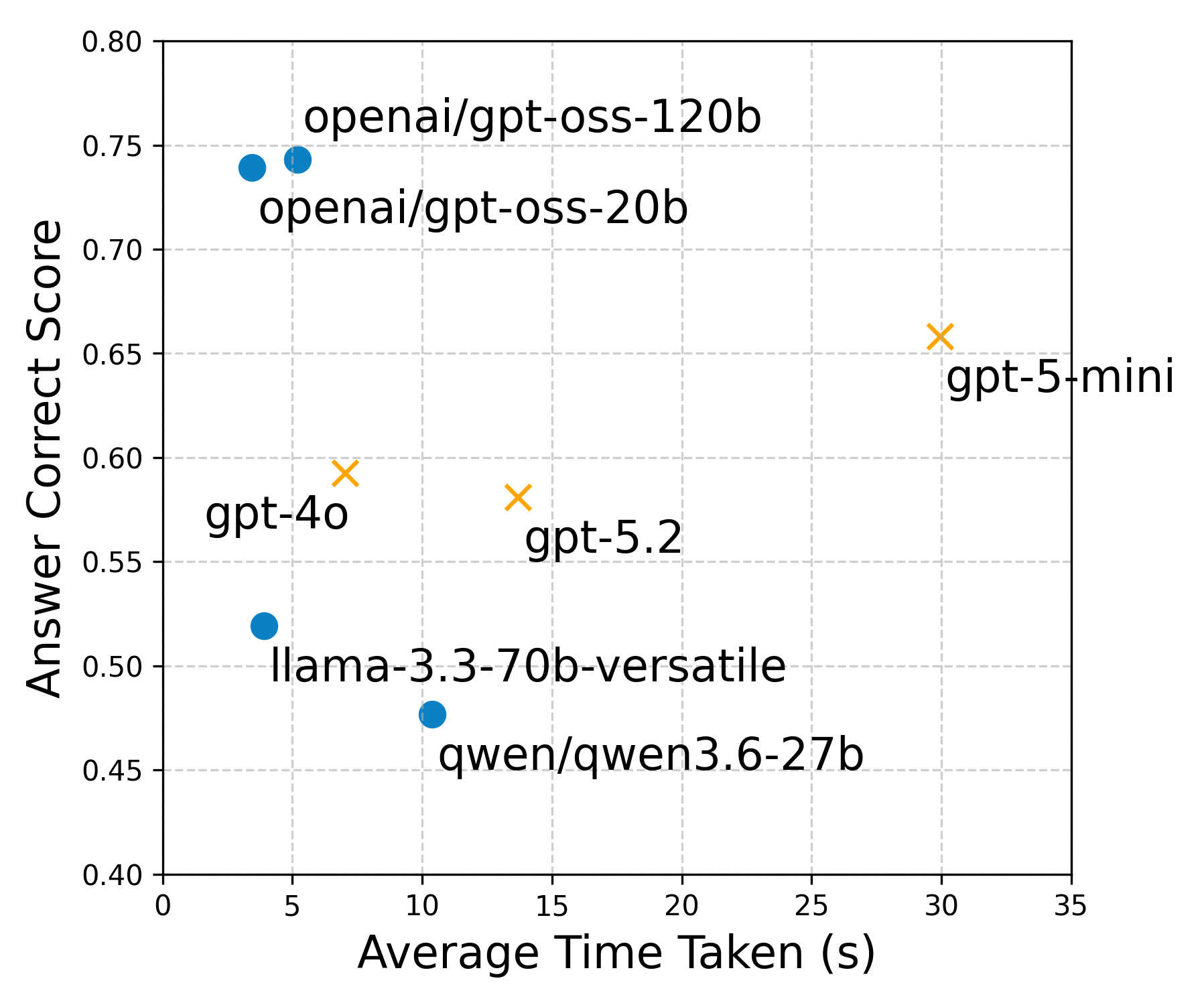}
    \caption{Comparison of LLM answer accuracy and mean delivery time. Groq API models are shown in blue, while OpenAI API calls are shown in orange. The distribution of per-call response times is shown in \cref{app:llm_timing_variation}.}
\label{fig:llm_speed}
\end{figure}

The generation model was chosen after comparing answer quality and latency across several candidate models. Early versions used OpenAI's \texttt{gpt-4o-mini}. Subsequent testing through the Groq API allowed open-weight models to be compared on both axes. \Cref{fig:llm_speed} shows this comparison for models available through the OpenAI and Groq APIs, together with the distribution of response times over individual calls. In this benchmark, the Groq-served open-weight \texttt{gpt-oss} models achieve the highest answer-correctness scores with substantially lower and less variable latency than the OpenAI API calls. The \texttt{gpt-oss-20b} model is particularly relevant operationally because it can run on edge devices with 16~GB of memory~\cite{gptoss}.
The answer correctness shown in \cref{fig:llm_speed} was evaluated with the \texttt{chATLAS\_Benchmark} (\cref{sec:evaluation}) using an LLM-as-judge metric. For each benchmark question an independent judge model (OpenAI \texttt{gpt-4o-mini}, at zero temperature) is shown the question, the reference answer, and the generated answer and asked, with a binary yes/no instruction, whether the generated answer correctly answers the question. The correctness score is the fraction of questions judged correct. The full retrieval pipeline is held fixed so that differences reflect the generator alone.

The models were evaluated using the standard reasoning configurations supplied by their respective APIs. The setting was \texttt{medium} for \texttt{gpt-oss-120b}, \texttt{gpt-oss-20b}, and \texttt{gpt-5-mini}; \texttt{default}, meaning reasoning enabled, for \texttt{qwen3.6-27b}; and \texttt{none} for \texttt{gpt-5.2}. Reasoning effort is not applicable to \texttt{llama-3.3-70b-versatile} or \texttt{gpt-4o}, which are not reasoning models. The calls were timed end-to-end, so the latency values in \cref{fig:llm_speed} are descriptive of user wait time rather than model inference speed.

These studies, together with considerations of cost and availability, shaped the current provider-agnostic configuration, which can use Groq-served open-weight models such as \texttt{gpt-oss-120b}, \texttt{gpt-oss-20b}, and \texttt{llama-3.1-8b-instant}, as well as hosted OpenAI models such as \texttt{gpt-5-mini}. Cost also differs substantially across providers. At the prices used for this evaluation, \texttt{gpt-oss-20b} via Groq cost US\$0.075 per million input tokens and US\$0.30 per million output tokens, compared with US\$0.25 and US\$2.00 respectively for OpenAI \texttt{gpt-5-mini}. Auxiliary tasks use smaller or more local models. Query contextualisation runs on OpenAI's \texttt{gpt-5-nano}, while the intent-routing and safety layer runs on \texttt{gpt-oss-20b} served through CERN's LiteLLM gateway (\cref{sec:routing}). No fine-tuning of any generative model was performed. Domain adaptation in \ours is confined to the embedding model, and the generators are treated as interchangeable, off-the-shelf components that can be replaced as stronger or more local options become available.

\subsection{Query Routing and Safety}
\label{sec:routing}
Every incoming query is first processed by a lightweight \emph{intent router} before any retrieval or generation takes place. The router is a single LLM call, using the same CERN-hosted \texttt{gpt-oss-20b}~\cite{gptoss} model served through the LiteLLM gateway at zero temperature. It classifies the request into one of a small set of intents and returns a structured decision. Depending on the intent, the system either proceeds with the normal retrieval-augmented pipeline or returns a fixed, retrieval-free response:
\begin{itemize}
    \item \textbf{ATLAS or HEP question:} the default case. The query continues to the hybrid retrieval and generation pipeline in the selected search or assistant mode.
    \item \textbf{Person lookup:} questions about who holds a given role are answered by directing the user to the authoritative ATLAS appointment database, rather than inferring the answer from documentation.
    \item \textbf{Dangerous request:} requests seeking actionable assistance for serious harm are refused, with a message reminding the user that use of \ours is governed by the CERN Code of Conduct.\footnote{\url{https://hr.web.cern.ch/cerns-values-and-code-conduct/}}
    \item \textbf{Out of scope or unclear:} queries unrelated to ATLAS, or too ambiguous to act on, receive a short explanation of the system's scope or a request for clarification.
\end{itemize}
Performing this step before retrieval keeps the safety decision independent of whichever documents happen to be retrieved, and avoids spending retrieval and generation effort on requests that should not be answered. The router is designed to fail open. If the routing call errors, the query proceeds through the normal pipeline so that availability is preserved. Each routing decision (the chosen intent, the backing model and prompt version, and a confidence score) is recorded in a dedicated database table, giving an audit trail and the data needed to monitor and refine the policy over time.

\subsection{Integrations}


To be useful in practice, \ours must connect with existing ATLAS workflows. The modular architecture allows it to act both as a callable service (via API) and as a consumer of other services (e.g.\ retrieving monitoring data). Integration with CERN PaaS (PostgreSQL with \texttt{pgvector}) ensures authentication and support within institutional infrastructure.
In addition to the web-based interface, \ours{} is integrated into the CERN Mattermost platform, allowing users to interact with the assistant directly within existing collaboration communication channels. In practice, an ATLAS member can ask a question in ATLAS Machine Learning Mattermost channel using the slash command \texttt{/chatlas <question>}.\footnote{Optional bracketed arguments can restrict the searched sources, set retrieval depths, or set a date filter, for example \texttt{/chatlas [twiki, mkdocs] [k=10] [k\_text=5] [01-01-2020] What is the crack veto?}.} Responses returned in Mattermost include both the generated answer and references to the underlying source documents. The Mattermost deployment shares the same backend services as the web interface.

\subsubsection{MCP Server}

To further extend \ours capabilities and enable integration with third-party AI agents, we have developed a Model Context Protocol (MCP) server. 
The Model Context Protocol, created by Anthropic, provides a standardised interface for connecting large language models with external tools and services~\cite{mcp}. Snappily described as ``the USB-C port for AI,'' MCP establishes a uniform way to connect LLMs to resources they need, functioning similarly to an API but specifically designed for LLM interactions. 
This protocol enables seamless scaling, simplified deployment, and asynchronous handling of requests, making it particularly valuable for integration with modern AI development environments such as GitHub Copilot in VSCode, AI chat apps (ChatGPT, Claude Desktop) or command-line coding agents (Claude Code, OpenCode, Gemini-CLI).

The MCP could also allow future \ours implementations to take advantage of \emph{agentic workflows} in which an LLM acting as an AI agent may perform repeated vector searches, in combination with additional retrieval tools or other sources of context, until it identifies suitable answers to complex and multi-faceted queries.

We implemented the server using FastMCP~\cite{fastmcp}, a Python framework that streamlines MCP server development with minimal boilerplate code by providing a high-level, Pythonic interface for building, managing, and interacting with MCP servers. 
The server is deployed on CERN's PaaS infrastructure, where MCP tools translate user prompts into vector searches against our knowledge base and return structured responses. 
Prompts and answers are exchanged with the server through streamable HTTP requests, enabling efficient real-time communication. 
This architecture scales well. The asynchronous nature of MCP tools enables parallel processing of multiple queries, while CERN PaaS can spawn additional server instances on demand to handle increased load.

%% file: sections/7.conclusions.tex
\section{Future Work and Conclusions}
\label{sec:conclusions}

Although already deployed for ATLAS users, \ours represents a first step toward embedding retrieval-augmented conversational systems into the everyday work of a large scientific collaboration. By combining hybrid retrieval, domain-adapted embeddings, and a reproducible benchmark, it provides source-attributed access to ATLAS documentation while remaining compatible with CERN authentication and data-governance requirements. Its modular design also provides a basis for richer, task-specific services, realising in modern form the vision of shared tools articulated in the 2003 ATLAS collaborative-tools white paper~\cite{goldfarb2003atlas}.

Near-term priorities are sustainability and corpus coverage. Moving generation to open-source models available within the CERN environment~\cite{Belck2025SmallLM} would remove the current dependence on personal API accounts reimbursed through university grants. Expansion of the indexed corpus will be coordinated with the new ATLAS documentation search to align source coverage, metadata, and update mechanisms while avoiding duplicated effort. Candidate additions include a broader selection of Indico slides, subtitles and transcripts from CERN video records, and structured information from systems such as Glance.

A conversational interface also offers a low-cost route to multilingual access. Exploratory use in French and Japanese indicates that the generator handles language switching well, but the primarily English embedding model has not been systematically evaluated for cross-lingual retrieval. Future work will therefore test query translation and multilingual embedding models, in coordination with the ATLAS Equity, Diversity \& Inclusion Office.

Stronger reasoning models also make iterative, agentic search a practical next step. Models available through the CERN AI Gateway~\cite{PinheiroPereira:2026aigw} could issue repeated searches and tool calls over the \ours retrieval layer for complex questions, although the additional inference cost and latency favour their use when adequate resources are available~\cite{gao2025ragreasoning}. The existing \emph{deep agents} extension provides an alpha testbed for this approach. Through its MCP interface, \ours can also expose retrieval to external agents and invoke domain tools such as ROOT MCP~\cite{rootmcp}; tighter links to code repositories, configuration databases, and analysis tooling would extend this model to code-aware assistance. Production feedback, although currently sparse, can guide evaluation and future tuning. The breadth of activity at the 2026 \emph{Knowledge Bases and Agents for CERN} workshop~\cite{CERNKBAgents2026} illustrates the growing community interest and rapid evolution of the available models and infrastructure, reinforcing the value of a system that can adopt new services without rebuilding its knowledge base.

Finally, developing \ours gives students transferable skills in software engineering, data pipelines, information retrieval, and applied language models. This is especially valuable because most physics students pursue careers outside academia~\cite{royalsociety2010}. Such work benefits both the collaboration and the wider scientific workforce, while shared retrieval, evaluation, access control, and modular interfaces provide durable infrastructure for increasingly capable AI-assisted scientific work.

%% file: sections/8.appendix.tex
\section{Example Interactions}
\label{app:examples}

This appendix shows a representative selection of real user queries together with the response \ours generated. Most are drawn from interactions that received explicit thumbs-up or thumbs-down feedback (marked accordingly). Since such votes are sparse, these are complemented by examples from the full production query log, shown without a vote marker, chosen to illustrate usage types the voted sample does not capture. The examples are presented without names, usernames, or other user identifiers. For brevity each response is given by its opening followed by ``\dots'' and a short note of the points it covered. These examples illustrate both the breadth of questions the system handles and its current limitations.

All examples shown are assistant-mode interactions. Roughly one third of logged queries instead use the retrieval-only search mode, typically as short keyword lookups (for example ``EMPFlow'' or ``mc20 mc23 difference''). Search mode was the interface default for much of the deployment period, and its share fell sharply once assistant mode became the default, so the mode split reflects interface defaults as much as user preference.

\subsection{The same question, two votes}
The two queries below ask essentially the same thing (whether calorimeter cluster coordinates are referenced to the detector origin or to a primary vertex), phrased slightly differently. Both received a careful, correctly hedged answer, yet one was up-voted and the other down-voted.

\paragraph{``Are the cluster coordinates measured with respect to detector origin, or with respect to some primary vertex?'' (up-voted).}
\ours replies that ``It depends on the cluster variant you read\dots''. It then explains that low-level cluster coordinates are defined in the detector frame while many physics-level quantities (topo-clusters for jets, calibrated photon variables) are vertex-corrected, that trigger objects store several coordinate flavours, and gives a practical checklist for identifying which convention a given branch uses, with cited sources.

\paragraph{``Are the cluster's coordinates given wrt detector origin, or are they given wrt some chosen PV?'' (down-voted).}
\ours gives a closely related answer, focused on the trigger/HLT case, explaining that quantities such as the seed coordinates are one of several stored position variants, that per-sampling coordinates may be absent in some derivations, and again offering a diagnostic checklist.

\noindent\emph{Comment:} two near-identical questions received comparably good answers but opposite votes. This illustrates the variability of generated responses and the subjectivity of a single-click feedback signal.

\subsection{Where it performs well}
The examples below span physics and analysis concepts, software and computing tasks, collaboration knowledge, and multilingual use, illustrating the breadth of questions \ours handles well.

\subsubsection{Physics and analysis concepts}

\paragraph{``How does ATLAS do b-tagging?'' (up-voted).}
\ours replies that ``ATLAS performs b-tagging with a chained flavour-tagging workflow\dots''. It then covers the low-level taggers (impact-parameter, secondary-vertex, track-based), the high-level neural-network taggers, modern transformer all-in-one taggers, where the chain runs (reconstruction and derivation), and the associated validation and monitoring tools, while noting that the algorithmic internals were not present in the retrieved context.

\paragraph{``What is the event weight? What are its units or type?'' (up-voted).}
\ours replies that ``the event weight is a scalar factor that multiplies the contribution of a Monte-Carlo event\dots''. It then explains what it encodes (generator cross-section, luminosity, pile-up reweighting and scale factors), that it is stored as a floating-point number in the MC event-weight vector, and that it is dimensionless, with supporting excerpts.

\paragraph{``What is a workspace?'' (up-voted).}
\ours replies that ``A workspace\dots is a ROOT file that contains a RooFit workspace\dots''. It then covers what it bundles (probability density functions, datasets, physics and nuisance parameters, configuration metadata), the tools that consume it, and how it is produced and used for statistical inference.

\paragraph{``What is the crack veto for electron reconstruction?'' (up-voted).}
\ours replies that the crack veto removes electrons reconstructed in the poorly instrumented barrel--endcap calorimeter transition, giving the standard excluded region $1.37 < |\eta| < 1.52$ and explaining that it improves the purity of the electron sample at the cost of some efficiency, with cited sources.

\subsubsection{Software and computing}

\paragraph{``What container should I use if I don't have CVMFS installed?'' (up-voted).}
\ours replies that the user should ``use a container image that already provides CVMFS inside the image\dots''. It then points to the relevant ATLAS and SFT image families, with the correct caveat that some images still expect CVMFS on the host and so the bundled-CVMFS images should be preferred.

\paragraph{``What script do I use to find out what is in an AMI tag?'' (up-voted).}
\ours replies that the AMI command-line client should be used and gives the exact command (\texttt{ami cmd GetAMITagInfo}) together with how to set up the ATLAS environment and the metadata fields it returns (description, the production r- and c-tags, input datasets, and output formats), citing the relevant documentation.

\paragraph{``How do I automatically resubmit SLURM \texttt{sbatch} array jobs?'' (up-voted).}
\ours notes that the documentation contains no dedicated recipe, then shows how to combine array submission with SLURM's requeue mechanism. The answer gives a batch script using the \texttt{-{}-requeue} directive, monitoring with \texttt{squeue}, an optional cap on the number of retries, and an alternative wrapper script that resubmits only the failed array indices. It transparently flags that this extrapolates beyond the retrieved sources.

\paragraph{``How to fix this error: \texttt{xAOD::shallowCopyContainer ERROR Couldn't set up DataLink to the original container's auxiliary store}''.}
\ours identifies the error as a problem with the auxiliary store of the original container and walks through the common causes (an uninitialised or missing auxiliary store, containers read without their store, copies made before the store is attached) with a corrective step for each. Pasting raw error output is a common usage pattern: the log contains many queries consisting largely of \texttt{Reco\_tf} tracebacks, grid-job failures, or TDAQ logs, for which the system typically identifies the failing component and proposes concrete debugging steps.

\paragraph{``How can I print the impact parameters ($z_0$ and $d_0$) of an \texttt{xAOD::Jet}?'' (up-voted).}
\ours explains that a jet carries a reference to its associated origin vertex, then returns a short, self-contained C++ example that loops over a jet container, retrieves the vertex via \texttt{originVertex()}, and prints the longitudinal and transverse quantities, with the important caveat that these jet-level values differ from the per-track impact parameters, and with pointers to the relevant software tutorials. The snippet is drawn from the existing tutorials rather than written from scratch, consistent with \ours{} acting as a documentation assistant rather than a code-generation tool.

\subsubsection{Collaboration knowledge and live data}

\paragraph{``How do I know when to use Internal, Preliminary or Work In Progress on my ATLAS plots?'' (up-voted).}
\ours replies with a table mapping each label to its use: \emph{Internal} for material still circulating for approval, \emph{Preliminary} for approved but unpublished results, \emph{ATLAS} for published results, and \emph{Work In Progress} for student talks, followed by a decision checklist and the governing policy sources. This is an example of institutional knowledge that is otherwise scattered across policy pages.

\paragraph{``Where can I find the total luminosity collected so far in Run~3?'' (up-voted).}
\ours directs the user to the ATLAS public luminosity results pages rather than quoting a specific number. Pointing to the authoritative, continuously updated source is the appropriate behaviour for a fast-changing quantity.

\paragraph{``How do I book shifts?''.}
\ours gives a step-by-step walkthrough of the OTP tool, including the correct URL, where to find the shift-booking interface, and how to select the relevant task. Questions about collaboration procedures (booking shifts, OTP accounting, qualification-task approval) are among the most common in the production log, reflecting the system's value for newcomers navigating collaboration processes.

\paragraph{``Who is the ID trigger coordinator?''.}
\ours answers that the coordinator is not identified in the retrieved documents and instead points to the trigger group mandate document and the relevant mailing lists. Declining to guess a name and deferring to the authoritative source is the appropriate behaviour for role-holder queries, which change over time. Such person-lookup queries are now also handled explicitly by the intent-routing layer (\cref{sec:routing}).

\subsubsection{Multilingual queries}

\paragraph{``Comment sont calibrés les électrons~?'' (``How are electrons calibrated?'', asked in French).}
\ours replies in French, outlining how electron selections are calibrated through data-to-simulation scale factors and the associated efficiency measurements, with cited sources. The production log also contains extended research dialogues conducted entirely in other languages, including a multi-turn investigation of large-R jet taggers in Japanese, answered with the relevant tagger recommendations and configuration details, and a multi-turn exchange in Chinese on configuring the IVF and AMVF vertexing algorithms. These support the multilingual observations discussed in \cref{sec:conclusions}.

\subsection{Limitations and open needs}

\paragraph{``What is `acceptance turn-on'?'' (down-voted).}
\ours replies that ``acceptance turn-on is the mass- or kinematic-dependent behaviour by which the analysis acceptance rises from zero to its plateau\dots''. It then explains why it matters (spectrum distortion, fit-range and selection choices) and how to characterise it. In our assessment this answer is accurate and useful despite the negative vote. It is a reminder that the sparse feedback signal can reflect rater subjectivity rather than answer quality.

\paragraph{``What is the meaning of trigger type 0x84?'' (down-voted).}
\ours returns a confident but unreliable answer associating the code with a specific front-end chip operation. The retrieved sources do not actually pin down the meaning. This is an example of highly specialised information that remains hard to find. It could be a genuine gap in corpus coverage rather than a phrasing problem.

\paragraph{``How many ATLAS Weeks have we had so far?'' (down-voted).}
\ours answers that there have been only two ATLAS (Overview) Weeks, in 2001 and 2002, which is incorrect. It matched a pair of old wiki pages instead of counting the hundreds of ATLAS Weeks that are catalogued as Indico events. Answering this correctly requires a live interface that can query and count Indico events, which \ours does not yet have (cf.\ the planned expansion of Indico coverage in \cref{sec:conclusions}).

\clearpage
\section{Details on benchmark choices}
\label{app:details_benchmark}
\subsection{Number of train and test (evaluation) questions}

\begin{figure}[htbp]
\centering
\begin{subfigure}{0.48\linewidth}
    \centering
    \includegraphics[width=\linewidth]{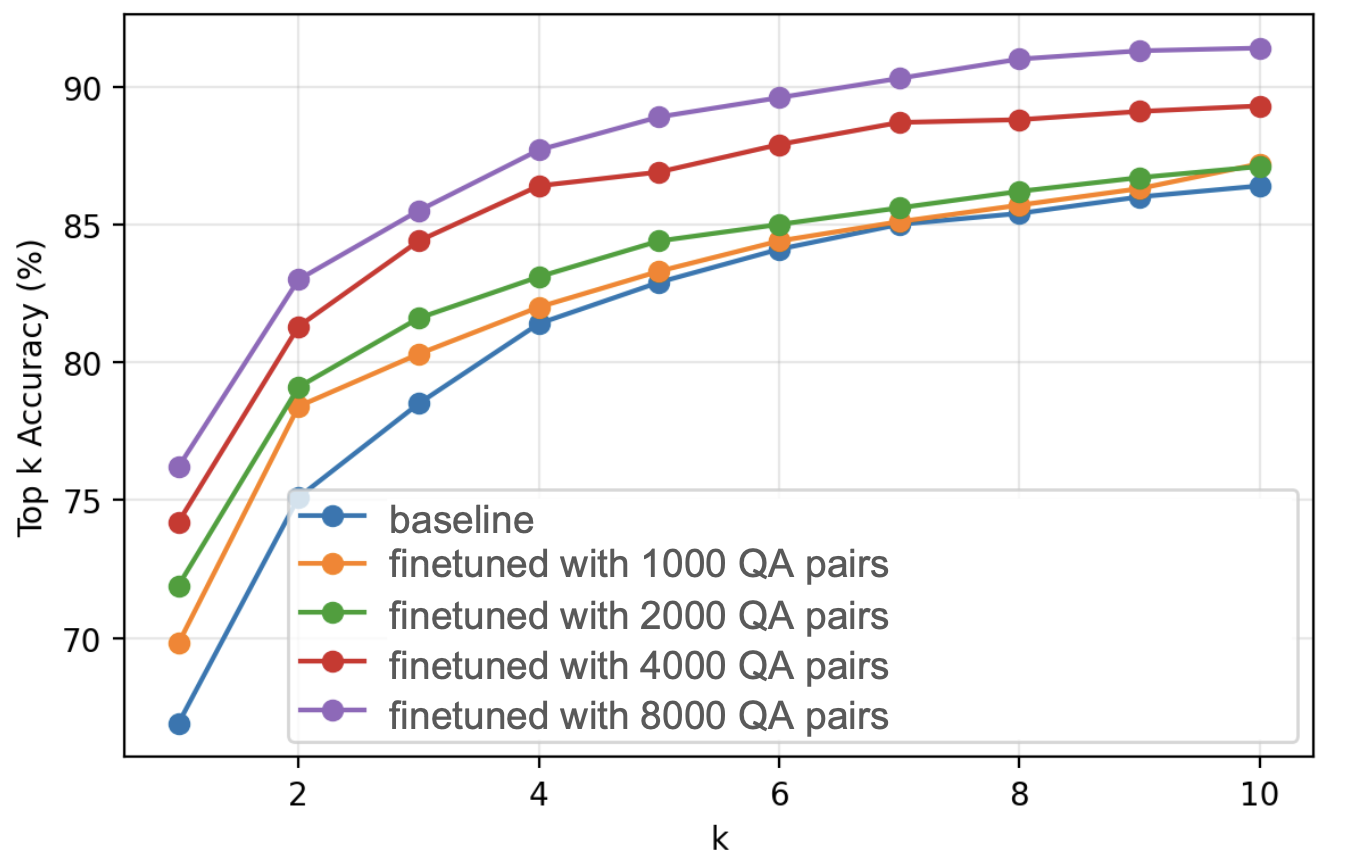}
    \caption{Evaluated on the GitLab benchmark}
    \label{fig:embedding_scaling_gitlab}
\end{subfigure}
\hfill
\begin{subfigure}{0.48\linewidth}
    \centering
    \includegraphics[width=\linewidth]{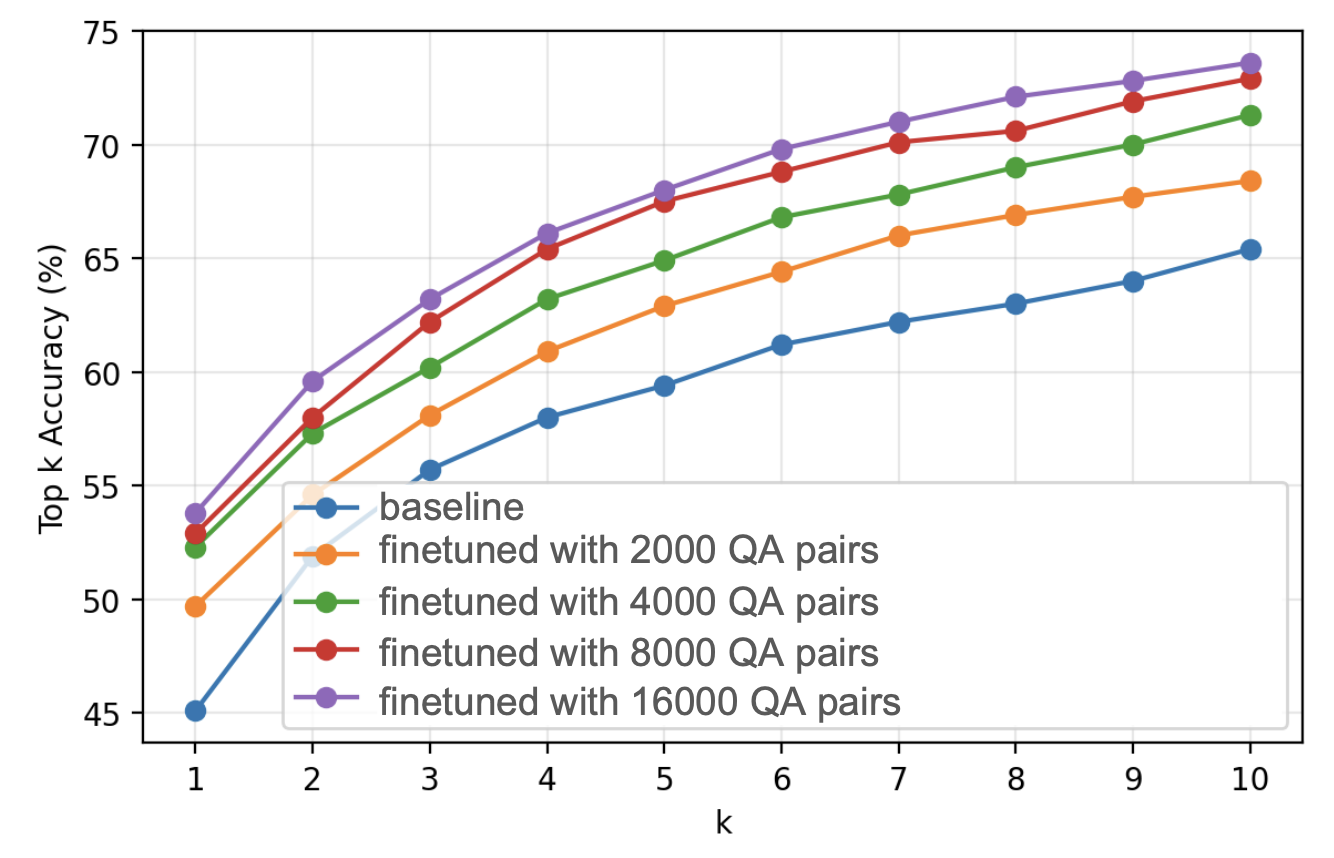}
    \caption{Evaluated on the ATLAS Talk benchmark}
    \label{fig:embedding_scaling_atlastalk}
\end{subfigure}

\caption{Retrieval performance as a function of fine-tuning dataset size. Top-$k$ retrieval accuracy is shown for embedding models fine-tuned using increasing numbers of question--answer (QA) pairs together with the untuned baseline. Both benchmarks exhibit substantial improvements over the baseline, with the largest gains achieved using the first few thousand QA pairs. Additional training data continue to improve retrieval accuracy but with diminishing returns, motivating our choice of 4,000 QA pairs per documentation source for the controlled cross-source evaluation.}
\label{fig:embedding_scaling}

\end{figure}

\Cref{fig:embedding_scaling} presents the effect of fine-tuning dataset size on retrieval performance when evaluated on the GitLab and ATLAS Talk benchmarks. In both cases, increasing the number of training QA pairs consistently improves top-$k$ retrieval accuracy over the untuned baseline, with the largest gains achieved using the first few thousand QA pairs. Additional training data continue to improve retrieval performance but with diminishing returns. While the GitLab benchmark exhibits gradual improvements across the full range of training set sizes considered, retrieval performance on the ATLAS Talk benchmark begins to plateau more rapidly. These scaling studies support our choice of 4,000 QA pairs per documentation source for the controlled cross-source evaluation, where equal-sized training sets were used to isolate the effect of the documentation source. In contrast, for the production embedding model, we fine-tune using all available QA pairs after reserving sufficient data for validation and evaluation, since maximising retrieval performance is preferred over maintaining equal training set sizes.

\section{Contextual Retrieval}
\label{appendix:contextual_retrieval}

This appendix provides the full definition of the marginal score used in \cref{sec:decisions}. The implementation of the contextual retrieval pipeline is available in the open-source \ours software packages~\cite{chatlas_embed,chatlas_chains}.

To quantify the separation between the correct parent document and its strongest incorrect competitor, we define the marginal score $M_i$ for query $q_i$ as
\begin{equation}
M_i = S_{i,\text{corr}} - S_{i,\text{incorr}},
\end{equation}
where
\begin{equation}
S_{i,\text{corr}}
=
\max_{c \in C(p_{i,\text{corr}})}
s(q_i,c),
\end{equation}
and
\begin{equation}
S_{i,\text{incorr}}
=
\max_{p \neq p_{i,\text{corr}}}
\max_{c \in C(p)}
s(q_i,c).
\end{equation}
Here, $q_i$ is the query, $p_{i,\text{corr}}$ is the correct parent document, $C(p)$ denotes the set of child chunks belonging to parent document $p$, and $s(q_i,c)$ is the cosine similarity score between query $q_i$ and chunk $c$. The quantity $S_{i,\text{corr}}$ represents the highest cosine similarity score among chunks belonging to the correct parent document, while $S_{i,\text{incorr}}$ represents the highest cosine similarity score among chunks belonging to the strongest competing incorrect parent document. A larger marginal score indicates that the retriever more strongly prefers the correct document over competing alternatives.

%

\clearpage
\section{LLM Timing Variation}
\label{app:llm_timing_variation}

\begin{figure}[H]
\centering
    \includegraphics[width=0.92\linewidth]{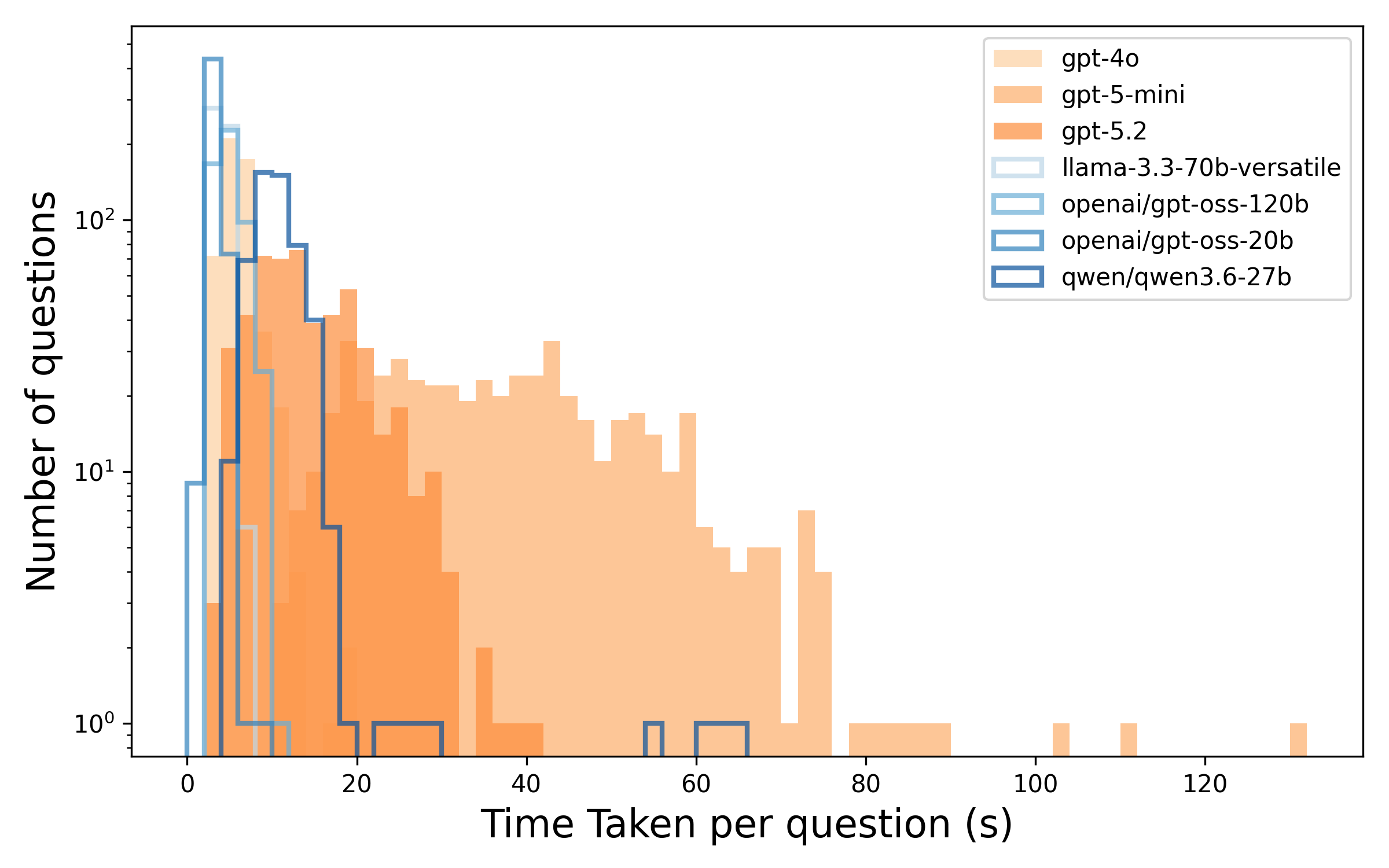}
    \caption{Distribution of response times across benchmark questions for OpenAI API calls (orange) and Groq API calls (blue).}
    \label{fig:llm_time_hist}
\end{figure}

\Cref{fig:llm_time_hist} shows the distribution of per-question response times for the model calls summarised by their mean latency in \cref{fig:llm_speed}. The OpenAI API calls are shown in orange and the Groq API calls are shown in blue. The long tail for the hosted OpenAI models is one of the practical considerations, alongside answer quality and cost, that motivates keeping the generator provider-agnostic.

